\documentclass[a4paper,10pt]{article}
\usepackage[pdftex]{graphicx}
\usepackage{a4wide}
\usepackage{amsmath, amsthm, amssymb}
\usepackage{multirow, amsfonts, mdwlist, wrapfig}
\usepackage{authblk} 
\usepackage{hyperref}

\usepackage{tikz}
\usetikzlibrary{shadows}
\usetikzlibrary{shapes,arrows,decorations.pathmorphing,backgrounds,fit,positioning}
\tikzstyle{every picture}+=[remember picture]

\usepackage[scaled]{helvet} %
\usepackage{courier} %
\usepackage{eulervm} %
\normalfont
\usepackage[T1]{fontenc}

\newcommand{\bit}{\begin{itemize*}}
\newcommand{\eit}{\end{itemize*}}
\newcommand{\matA}{\mathbf{A}}
\newcommand{\matB}{\mathbf{B}}
\newcommand{\matz}{\mathbf{0}}

\newcommand{\qe}{\textit{Q.E.}}
\newcommand{\qes}{\textit{Q.Es}}
\newcommand{\mat}[1]{ {\boldmath #1} }
\renewcommand{\mat}[1]{ \mathbf{#1} }

\newcommand{\eye}{\mathbb{I}}
\newcommand{\eig} { { \lambda _ 1 }}

\newcommand{\forTR}[1]{ {\em #1 }}
\renewcommand{\forTR}[1]{}
\newcommand{\hide}[1]{}
\newcommand{\tp} {tipping point}
\newcommand{\tps} {tipping point\  }
\newcommand{\s}{s}
\newcommand{\supermodel} {super-model}

\newcommand{\calJ}{{\cal J}}
\newcommand{\calD}{{\cal D}}
\newcommand{\lambdaJ}{\lambda_{\calJ}}
\newcommand{\lambdaA}{\lambda_{\matA}}
\newtheorem{informalTheorem}{Informal Theorem}
\newtheorem{theorem}{Theorem}
\newtheorem*{thm}{Theorem}
\newtheorem{lemma}{Lemma}
\newtheorem{assumption}{Assumption}
\newtheorem{observation}{Observation}
\newcommand{\vpm}{VPM}
\newcommand{\vect}[1]{ \mathbf{\vec{#1}} }
\newcommand{\sus}{\textbf{S}}
\renewcommand{\sus}{$S$}
\newcommand{\exps}{\textbf{E}}
\renewcommand{\exps}{$E$}
\newcommand{\infd}{\textbf{I}}
\renewcommand{\infd}{$I$}
\newcommand{\subscript}[1]{\ensuremath{_{\textrm{#1}}}}
\newcommand{\siiv}{S\textsuperscript{*}I\textsuperscript{2}V\textsuperscript{*}}

\begin{document}
\title{Got the Flu (or Mumps)? Check the Eigenvalue!}
\author[, 1]{B. Aditya Prakash\footnote{badityap@cs.cmu.edu}}
\author[, 2]{Deepayan Chakrabarti\footnote{deepay@yahoo-inc.com}} 
\author[, 3]{Michalis Faloutsos\footnote{michalis@cs.ucr.edu}}
\author[, 3]{Nicholas Valler\footnote{nvaller@cs.ucr.edu}}
\author[, 1]{Christos Faloutsos\footnote{christos@cs.cmu.edu}}
\affil[1]{Computer Science Department, Carnegie Mellon University, Pittsburgh}%
\affil[2]{Yahoo! Research, Sunnyvale}
\affil[3]{Department of Computer Science and Engineering, University of California - Riverside}

\date{March 30, 2010}
\maketitle

\begin{abstract}
For a given, arbitrary graph, 
what is the epidemic threshold?
That is, under what conditions will a virus result in an epidemic?
We provide the {\em \supermodel} theorem,
which generalizes older results
in two important, orthogonal dimensions. The theorem shows that
(a) for a wide range of virus propagation models (VPM) that include
{\em all} virus propagation models in standard literature
(say, \cite{hethcote2000}\cite{networksbook}),
and (b) for {\em any} contact graph,
the answer always depends on the first eigenvalue 
of the connectivity matrix.
We give the proof of the theorem, arithmetic examples
for popular VPMs, like flu (SIS), mumps (SIR), SIRS and more.
We also show the implications of our discovery:
easy (although sometimes \textit{counter-intuitive}) 
answers to `what-if' questions; 
easier design and evaluation of immunization policies, 
and significantly faster agent-based simulations.
\end{abstract}
\newpage
\section{Introduction - Preliminaries}

Given a social or computer network, 
where the links represent who has the potential to infect whom, 
can we tell whether a virus will create an epidemic, 
as opposed to quickly becoming extinct? 
This is a fundamental question in epidemiology; 
intuitively, the answer should depend on (a) the graph connectivity
and (b) the virus propagation model (VPM).
This threshold is the 
minimum level of virulence to prevent 
a virus from dying out quickly~\cite{kleinberg07}. 

No result till now has unified the varied observations 
on the effect of different network structures 
on the way different diseases spread. 
Moreover, with the exception
of our earlier work on the SIS model~\cite{deepay2008}
and its follow-up~\cite{ganesh2005};
no other analysis examines arbitrary-topology graphs:
the overwhelming majority of work
focuses either on full-clique topologies
(everybody contacts everybody else),
or on `homogeneous' graphs~\cite{kephart1991, kephart1993},
or on power-law graphs~\cite{vespignani2001}
or hierarchical (near-block-diagonal) topologies~\cite{hethcote1984}
(people within a community contact all others
in this community, with a few cross-community contacts).

We show that, irrespective of the virus propagation model, 
the effect of the underlying topology can be captured 
by just one parameter: 
the first eigenvalue $\eig$ of the adjacency matrix $\matA$. 
In particular we cover {\em all} models given 
in the standard survey by Hethcote~\cite{hethcote2000}, which includes models like SIS
(no immunity, like flu - `susceptible, infected,
susceptible') and SIR 
(life-time immunity, like mumps: `susceptible,
infected, recovered'). 
We also include numerous other cases like SIRS~\cite{networksbook} (temporary immunity), 
our own useful generalizations SIV (vigilance/vaccination
with temporary immunity) and SEIV 
(vigilance/vaccination with temporary immunity \textit{and} virus incubation) 
and many more. A few of these models are shown in Figure~\ref{fig:mh}, organized
in a lattice.

Informally, our result can be stated as:
\begin{informalTheorem}
    For \underline{any virus propagation model} (VPM)
    in the published literature,
    operating on an underlying undirected contact-network of \underline{any arbitrary topology} 
    with adjacency matrix $\matA$, 
    the epidemic threshold depends only on the first eigenvalue 
    \begin{equation*}
    \eig \nonumber
    \end{equation*}
    of $\matA$ and some constant $C_{\mathrm{\vpm}}$ that is determined 
    by the virus propagation model.
\end{informalTheorem}

Next, we give some preliminary definitions
and in the upcoming sections,
state our main result together with its application on a select few standard models and its potential uses (Section~\ref{sec:result-uses} and Section~\ref{sec:applications}) and finally give simulation results to illustrate the \supermodel\ theorem and discuss some of its implications (Section~\ref{sec:discussion}). The proof roadmap is explained in Section~\ref{sec:roadmap} with a description of the general model while the complete proof is presented in the Appendix.

\subsection{Preliminaries}
Table~\ref{tab:param} and Table~\ref{tab:models} list common terminology and describe some of the epidemic models we will be using in our article. Specific mathematical symbols and notations used for proofs are described in Appendix~\ref{sec:notation}.

\begin{table}[htb]
\caption{\textbf{Common Terminology}} \label{tab:param}
\centering
\begin{small}\begin{tabular}{||l|p{4in}||}
\hline \hline
\vpm\          & virus-propagation model \\ \hline 
NLDS          & non-linear discrete-time dynamical system \\ \hline\hline
$\beta$       & attack/transmission probability over a contact-link  \\ \hline
$\delta$      & healing probability once infected \\ \hline
$\gamma$  & immunization-loss probability once recovered (in SIRS) or vigilant (in SIV, SEIV) \\ \hline
$\epsilon$   & virus-maturation probability once exposed - hence, $1 - \epsilon$ is the virus-incubation probability \\ \hline
$\theta$      & direct-immunization probability when susceptible \\ \hline
$\matA$     & adjacency matrix of the underlying undirected contact-network \\ \hline
$N$            & number of nodes in the network \\ \hline
$\eig$         & largest (in magnitude) eigenvalue of $\matA$ \\ \hline 
$s$             & effective strength of a epidemic model on a graph with adjacency matrix $\matA$ \\ \hline
\end{tabular}\end{small}
\end{table}

\begin{table}[htb]
\caption{\textbf{Some Virus Propagation Models (\vpm s)}} \label{tab:models}
\centering
\begin{small}\begin{tabular}{||l|p{4in}||}
\hline \hline
SIS &  `susceptible, infected, susceptible' \vpm\ - no immunity, like flu  \\ \hline 
SIR & `susceptible, infected, recovered' \vpm\ - life-time immunity, like mumps \\ \hline 
SIRS & \vpm\ with temporary immunity \\ \hline 
SIV & `susceptible, infected, vigilant' \vpm\ - immunization/vigilance with temporary immunity \\ \hline 
SEIR & `susceptible, exposed, infected, recovered' \vpm\ - life-time immunity \textit{and} virus incubation \\ \hline
SEIV & \vpm\ with vigilance/immunization with temporary immunity \textit{and} virus incubation \\ \hline
\end{tabular}\end{small}
\end{table}

\begin{figure}[htbp]
\begin{centering}
\begin{tikzpicture}[scale=1]%
   \tikzstyle{every node}=[fill=blue!65,draw,circle,thick,text=white, text badly centered, circular drop shadow]
    \tikzstyle{edge from parent}=[draw=none]%

   \node (SIIV) {\begin{large}\textbf{\siiv}                 \end{large}} 
    child { child {node (RAND1) {$\ldots$}}}
    child { child {node (RAND2) {$\ldots$}}}
    child { child {
    child {  
    child {  
    child { node (MSEIV) {\begin{small}\textbf{MSEIV}\end{small}}
    	child {node[fill=brown] (MSEIR) {\begin{small}\textbf{MSEIR}\end{small}}}
    	child { 
		child { 
    		child {node (SEIV) {\textbf{SEIV}} 
     			child {node[fill=brown] (SEIR) {\textbf{SEIR}} }
     			child {
     			child {node (SIV) {\textbf{~~SIV~~}} 				
			    	child {node[fill=brown] (SIRS) {\textbf{SIRS}}
			        	child {node[fill=brown] (SIR) {\textbf{~~SIR~~}}}
	       	 		child {node[fill=brown] (SIS) {\textbf{~~SIS~~}} }
	    			}
    			}
			}
			child {node[fill=brown] (SEIS) {\textbf{SEIS}}}			
                      }}}
    	child {node[fill=brown] (MSEIS) {\begin{small}\textbf{MSEIS}    	                     \end{small}} } }
     }}}} 
    child { child {node (RAND3) {$\ldots$}}}
    child { child {node (RAND4) {$\ldots$}}};

\begin{scope}[dotted, thick]
\draw (MSEIR) -- (SEIR) ;
\draw (MSEIS) -- (SEIS) ;
\draw (MSEIV) --  (SEIV) ;
\end{scope}
\begin{scope}[loosely dotted, thick]
\draw (SIIV) --  (MSEIV)  ;
\draw (SIIV) -- (RAND1) ;
\draw (SIIV) -- (RAND2) ;
\draw (SIIV) -- (RAND3) ;
\draw (SIIV) -- (RAND4) ;
\end{scope}
\draw (MSEIV) --  (MSEIS) ;
\draw (MSEIV) --  (MSEIR) ;
\draw (SEIR) -- (SIR) ;
\draw (SEIS) -- (SIS) ;
\draw (SEIV) -- (SIV) ;
\draw (SEIV) -- (SEIR) ;
\draw (SEIV) -- (SEIS) ;
\draw (SIV) -- (SIRS) ;
\draw (SIR) -- (SIRS) ;
\draw (SIS) -- (SIRS) ;
\end{tikzpicture}
\caption{\textbf{Virus Propagation model hierarchy (actually, lattice) for some standard models including
SIRS (temporary immunity),
SIV (vigilance, i.e., pro-active vaccination); SEIV (includes
the `exposed but not infectious' state, and
temporary vigilance); MSEIR (with the passive immune state $M$); and our main generalization \siiv. The brown colored nodes denote standard \vpm s found in literature while the blue colored nodes denote our generalizations. Each \vpm\ is a generalization of all the models below it. \label{fig:mh}}}
\end{centering}
\end{figure}
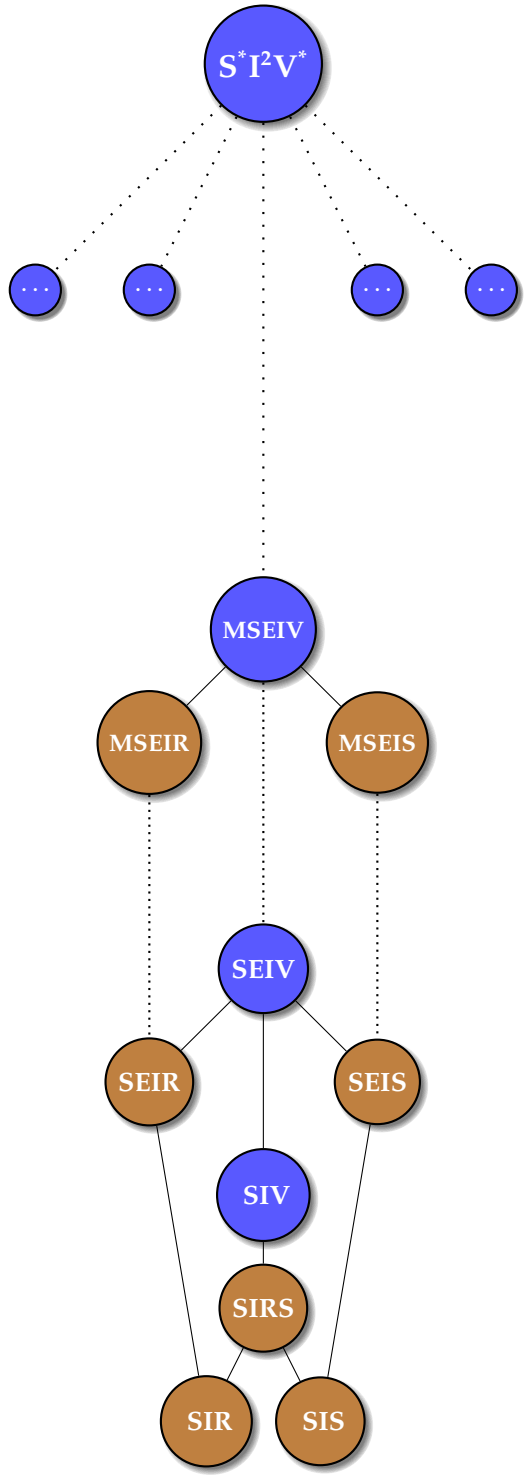

\typeout{importing a tikz drawing}
Figure~\ref{fig:mh} shows the generalization hierarchy for some common epidemic models. The brown colored nodes denote standard \vpm s found in literature while the blue colored nodes denote our generalizations. Each \vpm\ is a generalization of all the models below it e.g. SIV is a generalization of SIRS, SIR and SIS. Our main generalization, \siiv, is described in detail later in Section~\ref{sec:genmodel}.

\vspace{1in}
\section{Main Result}
\label{sec:result-uses}

The \tps for each of the models captures a fundamental
transition in the behavior of the system
between the two extremes:
a network-wide epidemic,
versus a minor local disturbance that fizzles out quickly.
We use the typical definition of the threshold used in literature~\cite{deepay2008, hethcote2000, barratbook, ganesh2005}. Intuitively, below threshold 
the virus has no realistic chance of spreading the infection while above threshold the virus can take over and create an epidemic. For the SIS model, the \tps describes
some maximum {\em strength} of a virus, that will
guarantee no epidemic~\cite{deepay2008, hethcote2000}. 
We define {\em strength} a little later (see Equation~\ref{eq:strength}).
Similarly for the SIR model, the \tps relates the explosiveness of the
infection phase w.r.t. the virus strength, since in this model the virus will become
extinct. 

In order to standardize the discussion of threshold results, we cast the threshold problem as expressing the normalized \textit{effective strength} of a virus as a function of the \textit{particular} propagation model and the \textit{particular} underlying contact-network. So we are `above threshold' when the effective strength $\s > 1$, `under threshold' when $\s < 1$ and the threshold or the \tps is reached when  $\s = 1$.

Intuitively, the effective strength $\s$ can be thought of as the basic reproduction number $R_0$~\cite{hethcote2000} frequently used in epidemiology. The effective strength is roughly the generalized $R_0$ for the virus model and an arbitrary graph and is the quantity which determines the \tps of an infection over a contact-network.\\\\\\

Formally, our main result is:

\begin{theorem}[Super-model theorem - sufficient condition for stability] \label{thm:main}
For virus propagation models which satisfy our general initial assumptions (see Section~\ref{sec:genmodel}) and for any arbitrary undirected graph with adjacency matrix $\matA$ and largest eigenvalue $\eig$, the sufficient condition for stability is given by:
\begin{equation}
\boxed{s  < 1}
\end{equation}
where, $s$ (the \textit{effective strength}) is:
\begin{equation}
\label{eq:strength}
\boxed{s = \eig \cdot C_{\mathrm{\vpm}}}
\end{equation}
\noindent
and $C_{\mathrm{\vpm}}$ is a constant dependent on the virus propagation model (given by Equation~\ref{eq:c} in the Appendix). Hence, the \tps is reached when $s = 1$.
\end{theorem}

\subsection{Two, Orthogonal Generalizations}
Our result generalizes along two different, difficult directions:
(a) arbitrary contact-network topologies and 
(b) several virus propagation models (\vpm s). 

\subsubsection{General Topologies}
Much of previous work~\cite{andersonmay} has concentrated 
on the analysis of \vpm s on \textit{specific} types 
of contact-networks, typically cliques or homogeneous graphs.
We include them all, as \textit{special} cases. Specifically
\begin{itemize*}
 \item Cliques, where every node contacts every other node. In that case,
       our result gives $\eig = N$, 
       where $N$ is the number of nodes in the graph
 \item Homogenous graphs, with fixed degree $d$ and 
       random Erd\"{o}s-R\'{e}nyi graphs 
       with expected degree $d$
       (e.g. see~\cite{kephart1991, kephart1993}).
       In all these cases we have $\eig = d$, and our theorem includes
       the previous results.
 \item Hierarchical (i.e., near-block-diagonal),  e.g.~\cite{hethcote1984}.
 \item Power-law (e.g.~\cite{vespignani2001})
\end{itemize*}

Theorem~\ref{thm:main} provides a simple and natural generalization 
of these results to arbitrary graphs. 
For example, previous results~\cite{vespignani2001} 
have shown that the epidemic threshold in case of scale-free (power-law) 
networks is vanishingly small as the size $N$ of the network increases. 
This is a corollary of our theorem:
When a power-law graph grows ($N \rightarrow \infty$),
the largest eigenvalue grows with the highest degree, which also
grows infinity, and thus the threshold approaches zero.

\subsubsection{General \vpm s}
We also generalize with respect to
various \vpm s which satisfy our very general assumptions 
(see Section~\ref{sec:genmodel}). 
We refer to our generalized model as \siiv, because it has an arbitrary 
number of susceptible states, two infectious/infected states,
and an arbitrary number of vigilant/vaccinated (= recovered) states. All the standard models (like see~\cite{hethcote2000},\cite{networksbook}) are simply \textit{special} cases of \siiv:
\begin{itemize*}
\item the typical flu model, SIS, is a special case. 
\item the typical mumps model, SIR, which corresponds to permanent immunity, 
      is a special case, with one state for each class - $S$ belongs to the Susceptible class, $I$ belongs to the Infected class and $R$ belongs to the Vigilant class.
\item the SIRS model (temporary immunity), similar to the SIR model
\item the SEIRS model (\cite{hethcote2000}, page 601) where
     the virus has an incubation period 
     (state 'E': exposed, but not infectious), and all other ingredients
     of the SIRS model (temporary immunity).
\end{itemize*}
and more like SIV, SEIV, MSEIV etc. which generalize some specific models. The generalization hierarchy is shown in Figure~\ref{fig:mh}. We elaborate more on the above in Section~\ref{sec:examples}.

We now give a brief summary of our threshold results 
(Table~\ref{tab:threshold}) by applying Theorem~\ref{thm:main} 
on some standard epidemic models. 
Note the effect of the contact-network in effective strength 
for \textit{each} model is captured solely by one parameter, 
$\eig$ the first eigenvalue of the adjacency matrix of the network. 
Again, our result is a general one and these models just highlight 
the ready applicability of the result to standard \vpm s in use.
\begin{table}[htb]
\caption{\textbf{Threshold results for some models.}}
 \begin{center}
\renewcommand{\arraystretch}{2.5}
\begin{tabular}{|p{2in}|c|c|}
\hline 
\textbf{Model} & \textbf{Effective Strength ($\s$)} & \textbf{Threshold (\tp)} \\
\hline 
SIS &  \begin{large}$\eig \cdot$\end{large} \begin{Large}$\left(\frac{\beta}{\delta}\right)$\end{Large} & \multirow{7}{*}{\begin{Large}$\s = 1$\end{Large}}\\
\cline{1-2} 
SIR &  \begin{large}$\eig \cdot$\end{large} \begin{Large}$\left( \frac{\beta}{\delta}\right)$\end{Large} & \\ 
\cline{1-2} 
SIRS& \begin{large}$\eig \cdot$\end{large} \begin{Large}$\left( \frac{\beta}{\delta}\right)$\end{Large}& \\ 
\cline{1-2} 
SIV & \begin{large}$\eig \cdot$\end{large} \begin{Large}$\left( \frac{\beta\gamma}{\delta(\gamma+\theta)}\right)$\end{Large} &\\
\cline{1-2} %
SEIR & \begin{large}$\eig \cdot$\end{large} \begin{Large}$\left( \frac{\beta}{\delta}\right)$\end{Large} & \\
\cline{1-2} %
SEIV & \begin{large}$\eig \cdot$\end{large} \begin{Large}$\left(\frac{\beta\gamma}{\delta(\gamma+\theta)}\right)$\end{Large} &\\
\cline{1-2} %
SI\subscript{1}I\subscript{2}V\subscript{1}V\subscript{2} (\textit{used to model the H.I.V. virus, e.g. see~\cite{andersonmay}}) & \begin{large}$\eig \cdot$\end{large} \begin{Large}$\left(\frac{\beta_1 v_2 + \beta_2  \epsilon}{v_2 (\epsilon+v_1)}\right)$\end{Large} &\\
\hline 
\end{tabular}
\end{center}
\label{tab:threshold}
\end{table}

\section{Applications}
\label{sec:applications}
The results in this article can be fundamental to numerous applications.
We describe a few important ones next.

\subsection{Fast answers to ``what-if'' questions and guiding policy}
The threshold results can help quickly determine the result of plausible situations. For example, what happens if the virus is twice as infectious (virulent)? Similarly, what happens when there is a weaker strain of the virus? Our results will help in determining whether there is a danger of the infection taking-off or not. Naturally then this can feed into policy decisions for controlling epidemics. Assuming some models for the underlying contact network (like scale-free, small-world, hierarchical etc.) we can estimate which nodes/classes should be quarantined or immunized first. Given the linear dependence on $\eig$, we want to immunize nodes (and hence remove them from the contact graph) which will drop the $\eig$ value the most so that the resultant infection becomes below threshold and dies out. For example, they may decide to immunize teachers and kindergarten children first to control the epidemic. In addition, they can impose restrictions on travel so as to not increase the $\eig$ and hence the effective strength for the virus. The above discussion also illustrates the generality of our result. Policy makers can assume \textit{any} graph model which captures the contact behavior of the population the best and still use our threshold result to guide policy. 

A lot of work has been done to show that immunizing high-degree nodes in scale-free networks is a good idea because of the vanishing threshold result~\cite{vespignani2001}. But significantly, just concentrating on high-degree nodes will \textit{miss} those low-degree nodes which are good ``bridges'' and hence can have a important influence on decreasing $\eig$ when immunized. Intuitively, how \textit{disparate} the groups are to which a node connects is also important in addition to how \textit{many} groups one is connected to. For example, a single common friend of only some sportspeople and movie stars can have a huge impact in the outbreak of a disease even if he/she knows only a few sportspeople and movie stars (while sportspeople and movie stars are themselves very tightly connected). 

We have been concentrating on biological viruses only. But various biological virus models have been used to model computer viruses as well~\cite{kleinberg07} e.g. \cite{hayashi03} introduced the SHIR model ('susceptible', 'hidden', 'infected', 'recoverable') to model computers under email attack. More so than the biological cases, it is easier to get the entire underlying network. Hence our threshold results can be precisely used to make the network more robust to malware and computer viruses by selectively ``removing'' nodes from the contact-network by immunizing them like installing a firewall on them etc. 

\subsection{Simplifying epidemiological simulations}
 Epidemiological simulations in general have several parameters and are computationally expensive to run. By distilling the impact of the network topology and also giving the exact threshold, we can greatly simplify these efforts. For example, parameters which do not affect the effective strength of the contagion need not be varied. In addition, regular topologies like cliques, block-diagonal matrices lend themselves to fast eigenvalue computations. This can be taken advantage of to quickly identify parameter spaces where simulations would be useful.

\subsection{Viral Marketing}
A variety of dynamic processes on graphs are modeled like epidemic spreading. In contrast to the biological viruses, conversely, we may actually want the \textit{spread} of a contagion as quickly as possible in some situations e.g. spread of a product or idea in a network of individuals. The Bass model~\cite{bass69} fits product adoption data using parameters for pricing and marketing effects ignoring topology and hence assuming that all adopters have equal probability of influencing non-adopters. A more refined picture using our result can be constructed of when a product gains massive adoption on a network (equivalent to an ``epidemic''). Marketers can also then decide which target groups should the product be more marketed to and how aggressive should the marketing be for an estimated network of individuals. 

\subsection{Software patch distribution and more}
Another application is efficient spreading of software patches over a computer network. The patches behave like computer worms~\cite{msr-worm} and can help defend against other malicious worms. We want to maximize the spreading of the patch over the network. Given full knowledge of the router-network involved, we can estimate how ``aggressive'' the patch-worm (say by increasing the number of probes for possible hosts before dying out) has to be to initiate an ``epidemic''. Our threshold results can also help determine the ``vulnerability'' and hence consequently the cost of not patching a part of the network say after one wave to ``worm-patching''.

Various epidemic models have also been used to model blog cascades. These models can be now applied to arbitrary graphs e.g. study propagation of memes through blogs~\cite{leskovec09}.

\section{Proof Roadmap}
\label{sec:roadmap}
\begin{figure}[htb]
\centering
    \begin{tabular}{ccc}
\begin{tikzpicture}
 \draw[line width=1mm] (-2,0) parabola[bend pos = 0.5] bend +(0,2) +(4,0);
\node[draw=red, circle,fill=white,very thick,minimum size=0.5cm] at (0, 2.3) {};
\end{tikzpicture}& 

\begin{tikzpicture}
 \draw[line width=1mm] (-2,2) parabola[bend pos = 0.5] bend +(0,-2) +(4,0);
\node[draw=red, circle,fill=white,very thick,minimum size=0.5cm] at (0, 0.3) {};
\end{tikzpicture}&       

\begin{tikzpicture}
 \draw[line width=1mm] (-2,0) -- (2, 0);
\node[draw=red, circle,fill=white,very thick,minimum size=0.5cm] at (0, 0.3) {};
\end{tikzpicture}\\
      (A) Unstable  &  (B) Stable & (C) Neutral (at threshold) \\
  \end{tabular}
\caption{\textbf{Different types of equilibria for a discrete-time NLDS: Unstable, Stable and Neutral. The equilibrium is neutral at threshold.}}
\label{fig:ndlsequi}
\end{figure}
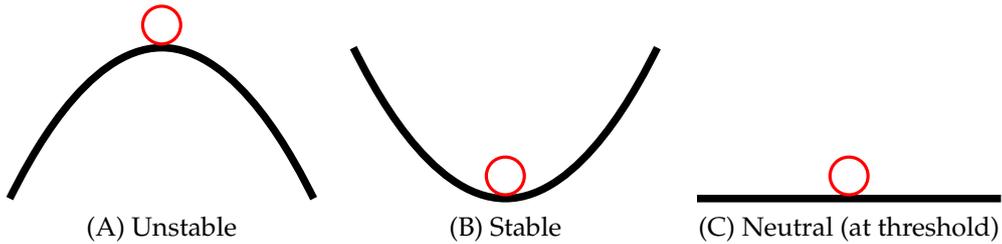

The first basic idea behind the proof is approximating 
the epidemiological models by a discrete time 
{\em non-linear dynamical system} (NLDS). 
A NLDS can be represented by \[\vect{P}_{t+1} = g(\vect{P}_t)\] 
where $g$ is some non-linear function operating on a vector. 
We define the vector $\vect{P}_t$ such that it specifies the state of the system at time $t$;
the exact definition will differ from model to model 
but it effectively encodes the probability of each node in the graph 
of being in any given state at time $t$. 
An equilibrium point (also called a fixed point) of the NLDS is the state vector (i.e. some particular $\vect{P}$) of the NLDS which does not change. Thus at the equilibrium point $\vect{P}_{t+1} = \vect{P}_t = \vect{x}$. Formulating a NLDS helps us in leveraging 
the vast results in NLDS literature. In particular, Theorem~\ref{theorem:asymstable} (given in Appendix~\ref{sec:syseqn}) gives the conditions when an equilibrium point a NLDS is stable (Figure~\ref{fig:ndlsequi} illustrates this concept). It relates the eigenvalues of the Jacobian of the NLDS at the equilibrium point with the stability of the NLDS at that point.

Intuitively, the \tps for any model then deals with analyzing 
the stability of the corresponding NLDS at the point 
when none of the nodes in the graph are infected, 
because otherwise the infection can still spread. If the equilibrium is unstable, a small ``perturbation'' (physically in the form of a few initial nodes getting infected) will push the system further away which physically means more and more nodes will get infected leading to an epidemic. But if the equilibrium is stable, the system will try to come back to the fixed point without going ``too-far'' away in effect, ``controlling the damage''. At threshold, the tendencies to go further away and come-back will be the same. In other words, the equilibrium is stable below the threshold and is neutral at the \tp.

The requirement imposed by Theorem~\ref{theorem:asymstable} on the eigenvalues of the `Jacobian' of the corresponding NLDS for any virus propagation model eventually reduces to a simple condition on the eigenvalue of the adjacency matrix. This condition translates into the effective strength of the virus under the model. The reason we can reduce the condition to one on the adjacency matrix is due to the special structure of the models, which is explained next. 

\subsection{General Model}
\label{sec:genmodel}
\begin{figure}[htb]
\begin{centering}
\begin{tikzpicture}[remember picture]
\node[cloud, draw, cloud puffs=20, aspect=1.75, cloud puff arc=140,fill=white,drop shadow] at (-4.5, 0) (sus)
	{\begin{tikzpicture}[place/.style={fill=blue!65,draw,circle,thick,text=white, text badly centered}]
	\node[place] at (-1, 0) (S1) {$S_1$}; 
	\node[place] at (0, 0) (S2) {$S_2$}; 
	\node at (1, 0) (S3) {$\ldots$}; 
	\node[rectangle]  at (0,1) {\textbf{`Susceptible'}};
	\end{tikzpicture}};
\node[starburst, starburst points=25, draw,fill=white,drop shadow] at (4.5, 0) (infd)
	{\begin{tikzpicture}[place/.style={fill=blue!65,draw,circle,thick,text=white, text badly centered}]
	\node[fill=blue!65,draw,circle, line width=1mm,text=white, text badly centered] at (-1, 0) (I1) {$I_1$}; 
	\node[place] at (0, 0) (I2) {$I_2$}; 
	\node at (1, 0) (I3) {$\ldots$}; 
	\node[rectangle]  at (0,1) {\textbf{`Infected'}};
	\end{tikzpicture}};
\node[ellipse, draw,fill=white,drop shadow] at (0, -4.5) (vigi)
	{\begin{tikzpicture}[place/.style={fill=blue!65,draw,circle,thick,text=white, text badly centered}]
	\node[place] at (-1, 0) (V1) {$V_1$}; 
	\node[place] at (0, 0) (V2) {$V_2$}; 
	\node at (1, 0) (V3) {$\ldots$}; 
	\node[rectangle]  at (0,1) {\textbf{`Vigilant'}};
	\end{tikzpicture}};
\end{tikzpicture}
\begin{tikzpicture}[remember picture, overlay]
\begin{scope}[>=stealth]
 \draw[red!80!black,line width=2mm,->,decorate,decoration={snake,amplitude=1.5mm,segment length=9mm,post length=3mm}]  
(sus) -- (I1) 
node[below=3pt,text=black,pos=0.4,text width=4cm,text centered] {Exogenous Transitions (depends on neighbors)}; 
\draw[green!60!black, line width=2mm, ->] (vigi.west) to [bend left=10] (sus.south)
node[left,text=black,midway,text width=3.7cm,text centered] {Endogenous Transitions};
\draw[green!30!black, line width=2mm, ->] (sus) to [bend left=10] (vigi.north west);
\draw[yellow!40!black, line width=2mm, ->] (infd) to [bend left=10] (vigi.east)
node[right,text=black,midway,text width=3.7cm,text centered] {Endogenous Transitions};
\draw[brown, line width=2mm, ->] (infd) to [bend right=30] (sus)
node[above,text=black,midway,text width=3.7cm,text centered] {Endogenous Transitions};
\end{scope}
\end{tikzpicture}
\caption{\textbf{General State Diagram for a node in the graph - it is \textit{not} a simple Markov chain. In essence, there are only three types of state `classes' - Susceptible, Infected and Vigilant. Only cross-class transitions have been shown. Note the unidirectional arrow from Infected to Vigilant. Only one type of transition is exogenous i.e. \textit{graph-based} (red curvy arrow) affected only by the neighbors of the node, all other transitions shown are endogenous. Any transitions between the states of the same class are also endogenous. The red arrow always ends at $I_1$, i.e. any state in the Infected class can cause a graph-based transition from any state in the Susceptible class only to the $I_1$ state.\label{fig:statedia}}}
\end{centering}
\end{figure}
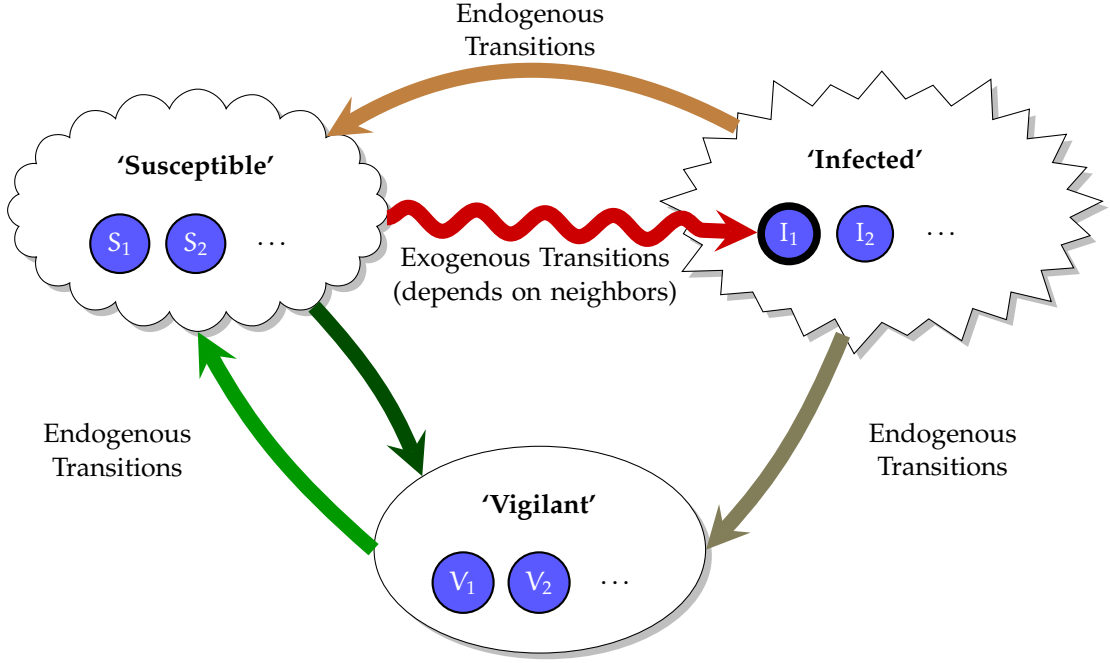

Suppose the virus-propagation model has $m$ states (e.g. $m=3$ for the SIR model with states \sus, \infd\ and $R$) and it operates on a graph of $N$ nodes. 
Consider then a $m\cdot N \times 1$ vector $\vect{P}_t$ 
which captures the probability of each node being in any of $m$ states
at a given time $t$.
Specifically:
\begin{equation}
\label{eq:probvec}
\vect{P}_t = \left[ \begin{matrix} 
                     P_{s_1,1,t} \\
		     P_{s_1,2,t} \\
		     \vdots \\
                     P_{s_1,N,t} \\
		     P_{s_2,1,t} \\
		     \vdots \\
		     P_{s_{m},N,t} \\
                    \end{matrix} \right]
\end{equation}

\noindent
where, $P_{s_j,i,t} = $ the probability of node $i$ 
being in state $s_j$ at time $t$. 
This vector completely defines the system at time $t$ 
and our NLDS equation will track the evolution of this vector across time.  

Every epidemic model will have some fundamental states 
and the choice of which states to include in a model depends on 
the particular disease characteristics~\cite{hethcote2000}. In fact, every model can be essentially thought of having states in any of the following broad classes (see Figure~\ref{fig:statedia}): 

\begin{description*}
 \item[\textit{Susceptible} Class:] Nodes in any of the states in this class signify that it (the individual) is  susceptible and can get infected by any neighboring node in a state of the Infected class. 

\item[\textit{Infected} Class:] In a state of this class, the node is infectious in the sense that it is capable of transmitting the infection to its neighbors. Note that each such state will have a \textit{transmissibility} parameter (e.g. $\beta$ in the SIR model for the infectious state \infd). This definition is general enough to accommodate models with states which have their transmissibility parameter = 0 i.e. they are `exposed' but not infectious (e.g. the \exps\ state in the SEIR model is a state which is in the Infected class in the sense that it can potentially cause infections but is not by itself infectious).

\item[\textit{Vigilant/Vaccinated} Class:] This class contains states which are not in either of the other two classes Susceptible and Infected. Nodes in any of the states in this class can not get infected nor do can they potentially cause infections. States like $M$ (called as the passive immune state which feeds to the susceptible state), $R$ (the recovered/died state where the node either gets permanent immunity or dies and hence doesn't participate in the epidemic further) etc. are all conceptually of the Vigilant type. Notice that importantly by definition, any state in the Vigilant class will not have a direct transition to any state in the Infected class. This is so because otherwise that state can be potentially infectious and thus is not part of the Vigilant class. 
\end{description*}

Apart from the types of states, models also fundamentally have only two types of transitions: \textit{Exogenous} (graph-based, in particular affected \textit{only} by the neighbors) and \textit{Endogenous} (caused by the node itself by some probability at every time step). For example, the transition from \sus\ to \infd\ in the SIRS model is an exogenous transition while the transition from $R$ to \sus\ is an endogenous transition. It is the presence of the graph-based transitions that makes our model \textit{not} a simple Markov chain and brings in the topology of the graph into play. In our general model note that only one class of transitions is graph-based, all others are endogenous transitions. The exogenous transitions are the ones that take a node from a state in the Susceptible class to the Infected class. In addition we assume any such exogenous transition ends at the $I_1$ state; but these transitions will themselves be caused by some neighbor node in any of the states of the Infected class. 

\begin{assumption}[Transition Assumption 1] The only way to get infected is through your neighbors i.e. there is no path \textit{to} a state in the Infected class from a state in the Susceptible class composed solely of endogenous transitions. 
\end{assumption}  
\begin{assumption}[Transition Assumption 2]
 Any exogenous (graph-based) transition always results in a transition from a state in the Susceptible class to the $I_1$ state. 
\end{assumption}

As seen from a single node, these models will look like in Figure~\ref{fig:statedia}. It gives the general state diagram for a node in the graph together with the assumptions on the transitions discussed above. We have shown only cross-class transitions and their types. All transitions between states of the same class are of course endogenous.

\subsection{Examples}
\label{sec:examples}
Our general characterization is powerful enough to seamlessly capture all the practical models like SIS, SIR, SIRS, SIV, SEIR, SERIS, MSIR, MSEIR etc.~\cite{hethcote2000} while still being mathematically tractable to yield simple threshold equations.

\begin{figure}[htbp]
\begin{centering}
\begin{tikzpicture}%
[scale=2,place/.style={fill=blue!65,draw,circle,thick,text=white, text badly centered,minimum size=1.2cm}]
   \node[place] at (-1, 0) (S) {\textbf{S}}; 
    \node[place] at (1, 0) (E) {\textbf{E}};
\node[place] at (2, -0.5) (I) {\textbf{I}};
\node[place] at (0,-1.5) (V) {\textbf{V}};
\draw[red,ultra thick,->,decorate,decoration={snake,amplitude=1mm,segment length=5mm,post length=1mm}] 
(S) -- (E) 
node[auto,midway,text=black] {$\beta$}; 
\draw[ultra thick, ->] (S) to [bend left=20] (V) 
node[auto,midway,text=black] {$\theta$};
\draw[ultra thick, ->] (V) to [bend left=20] (S) 
node[auto,midway,text=black] {$\gamma$};
\draw[ultra thick, ->] (E) to [bend left=10] (I) 
node[auto,midway,text=black] {$\epsilon$};
\draw[ultra thick, ->] (I) to [bend left=10] (V) 
node[auto,midway,text=black] {$\delta$};
\end{tikzpicture}
\caption{\textbf{Transition diagram for the SEIV model.}}
 \label{fig:seiv}
\end{centering}
\end{figure}
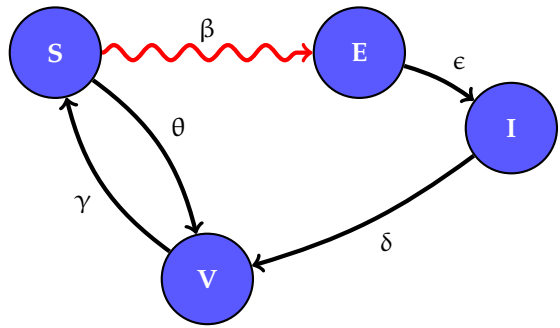

We give \textit{one} instantiation of our general model in the form of the SEIV model, which itself generalizes several known \vpm s (see Figure~\ref{fig:mh}). SEIV has one state (\sus) in the Susceptible class, two ($E$ and \infd) in the Infected class and one ($V$) state in the Vigilant class. The transition diagram for the SEIV model in shown in Figure~\ref{fig:seiv}. Note the similarities in Figures~\ref{fig:statedia} and \ref{fig:seiv} e.g. $E$ state is the $I_1$ state of Figure~\ref{fig:statedia}. The infection is caused only by the \infd\ state and the $E$ is only a latent state. The SEIR, SEIRS, SIRS, SIV, SIR and SIS models are all special cases of SEIV:
\begin{itemize*}
\item SIS, is a special case, with $\epsilon=1, \gamma=1, \theta=0$.
\item SIR, which corresponds to permanent immunity, with $\epsilon=1, \gamma=0, \theta=0$.
\item SIRS (temporary immunity), similar to the SIR model, with $\epsilon=1, \theta=0$.
\item SIV, our own model, where 'V' stands for vigilant or vaccinated, with $\epsilon=1$. 
\item SEIR, similar to SIR but where the virus has an incubation period, with $\gamma=0, \theta=0$.
\item SEIRS where the virus has an incubation period, and all other ingredients
     of the SIRS model, with $\theta=0$.
\end{itemize*}

\noindent
The complete proof for the general model with \textit{two} states in the infected class (we call this as the \siiv generalized model) is given in the Appendix. The proof follows the following roadmap:
\begin{enumerate*}
 \item We develop the system equations of our general NLDS in Appendix~\ref{sec:syseqn}.
\item The relevant fixed point is computed in Appendix~\ref{sec:fixedpt}.
\item The Jacobian at the fixed point is constructed in Appendix~\ref{sec:jacobian}.
\item Finally we prove Theorem~\ref{thm:main} by finding the condition to bound the eigenvalues of the Jacobian in Appendix~\ref{sec:eigenvalues} and Appendix~\ref{sec:stability}.
\end{enumerate*}

\forTR{
\begin{itemize}
 \item General Model
\item System Equations
\item Gradient Matrix
\item Threshold
\end{itemize}
}

\section{Discussion}
\label{sec:discussion}

We discuss some simulation examples for our threshold in some models and a few direct implications of the \supermodel\ theorem in this section. We also illustrate what the result implies for the `vulnerability' of the underlying contact graph for epidemics. Apart from the dependence of the threshold on $\eig$, it is instructive to note some unexpected results in specific models as well.

\subsection{Simulation Examples}
\label{sec:experiments}

\begin{figure}
\centering
    \begin{tabular}{cc}
	\includegraphics[width=3in]{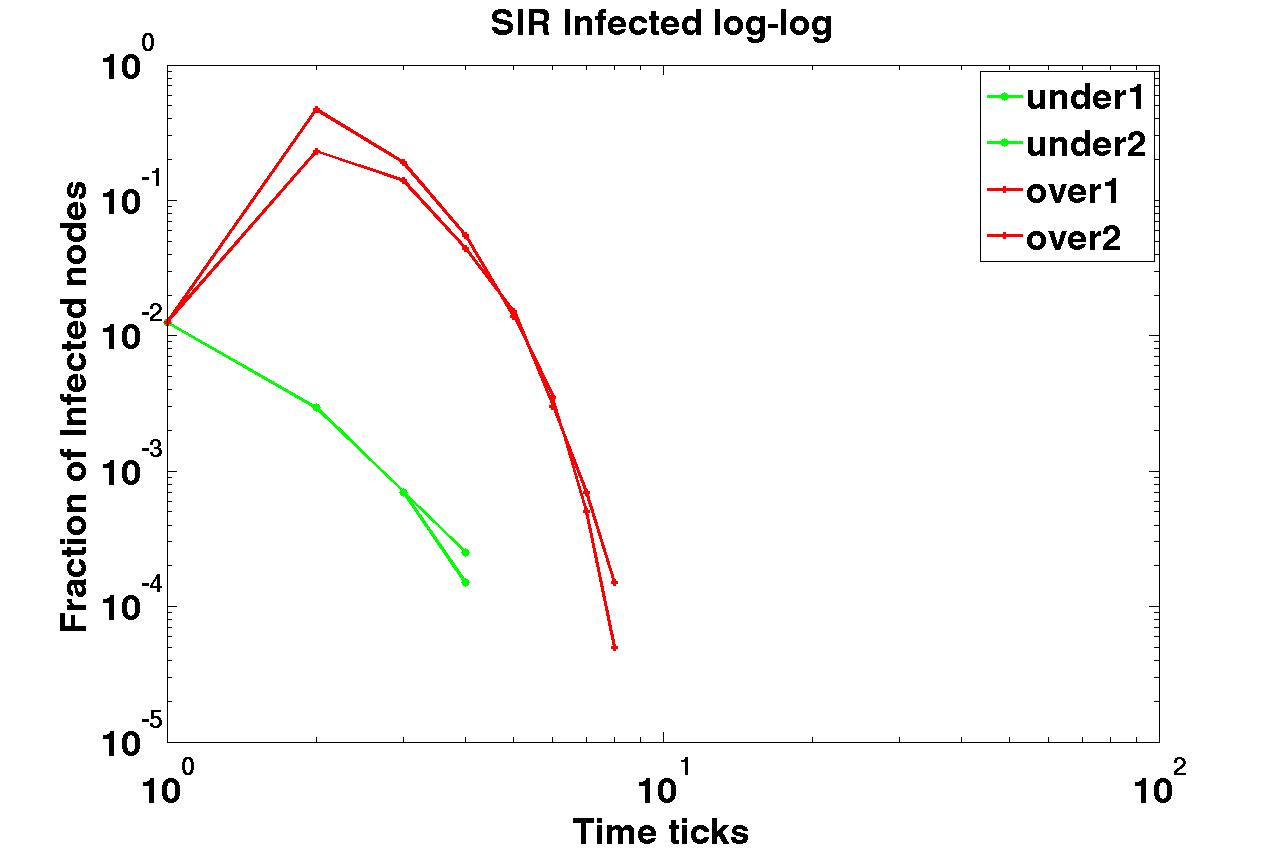} &
        \includegraphics[width=2.9in]{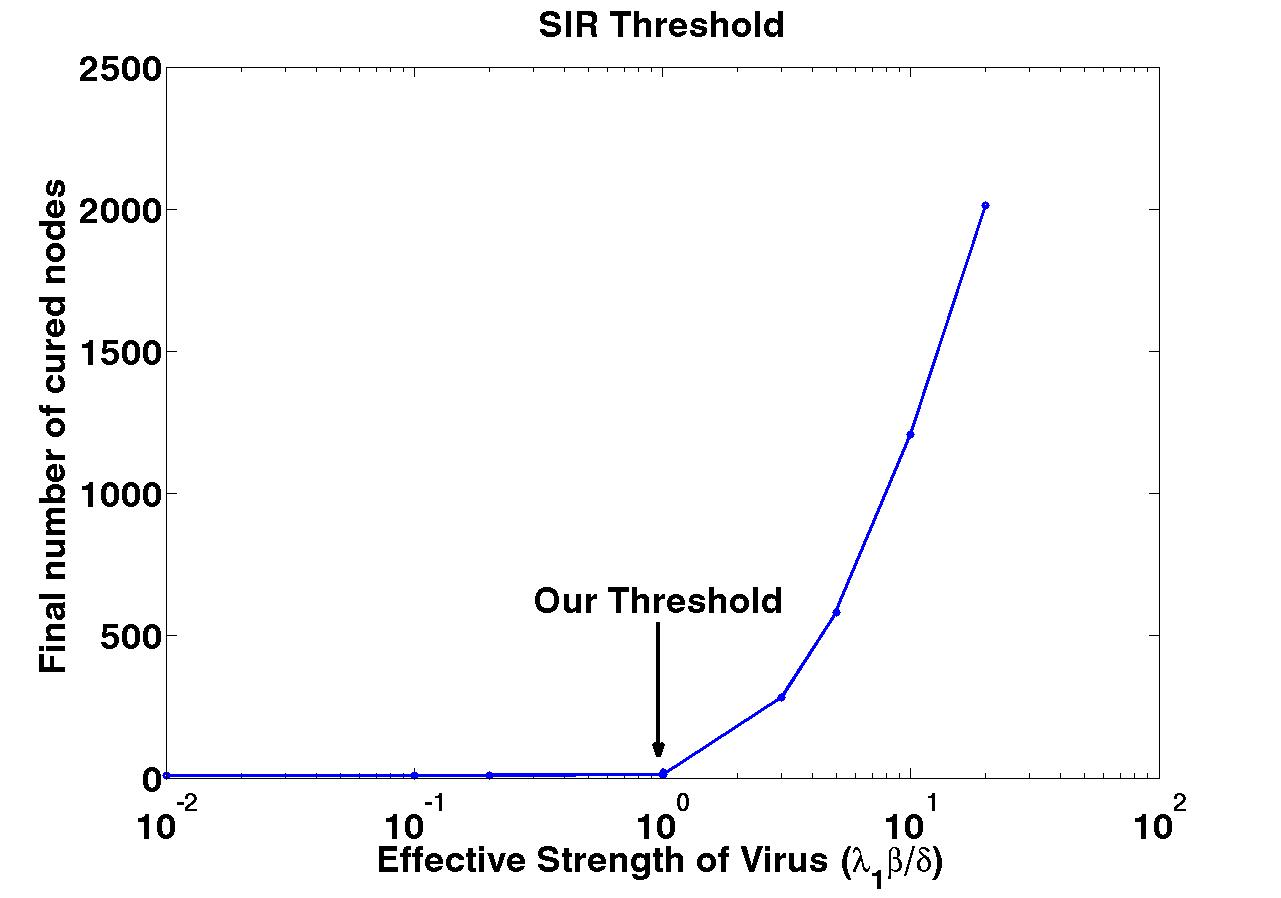} \\
	(A) SIR Infective Fraction Time Plot (log-log) & (B) SIR Footprint vs Strength (lin-log)  
  \end{tabular}
\caption{\textbf{SIR (all values averages over several runs): (A) Plot of Infective Fraction of Population vs Time (log-log). Note the qualitative difference in behavior \textit{under} and \textit{above} the threshold. (B) Plot of Final number of cured nodes (the \textit{footprint}) vs Effective Strength (lin-log). Note the \tps is exactly when the effective strength $s=1$.}}
\label{fig:infec-dynm-sir}
\end{figure}

\begin{figure}
\centering
    \begin{tabular}{cc}
	\includegraphics[width=3in]{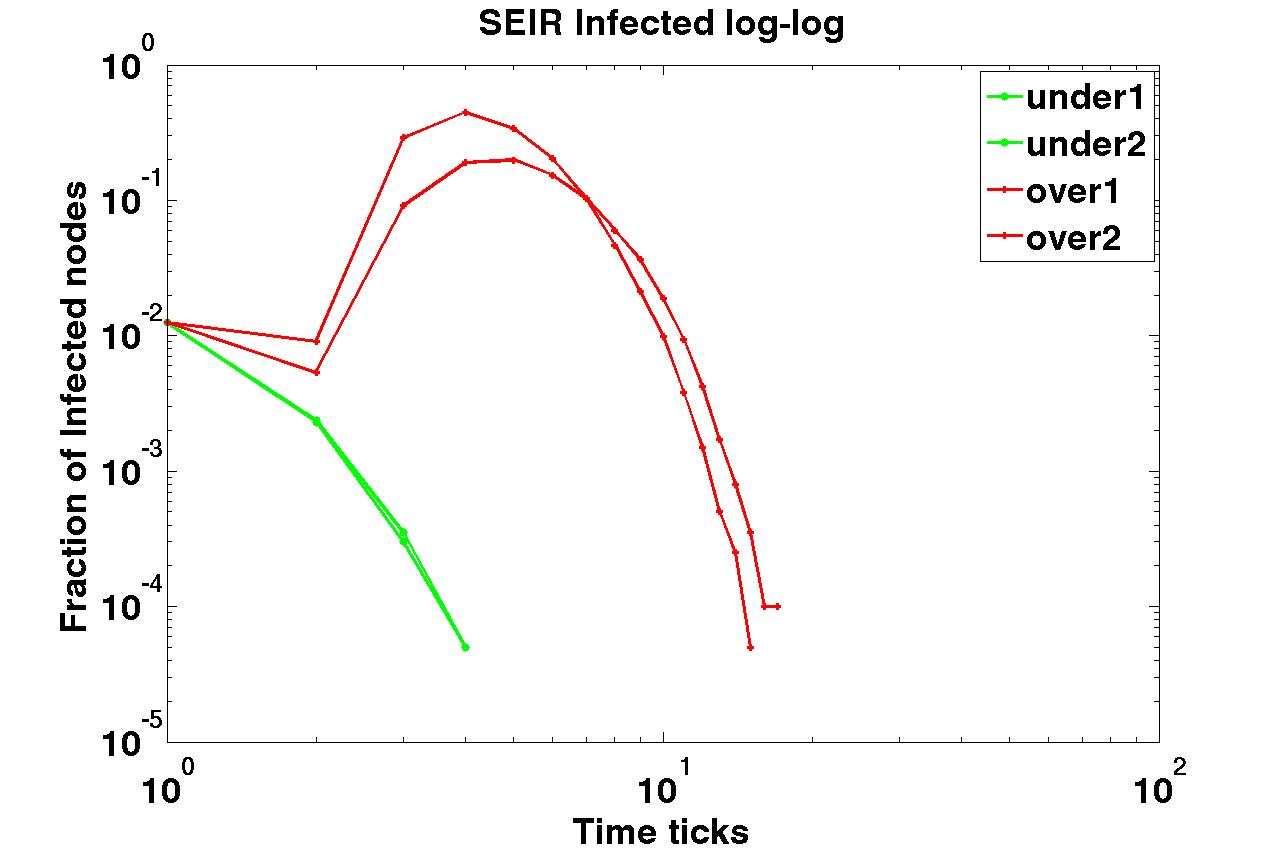} &
        \includegraphics[width=2.9in]{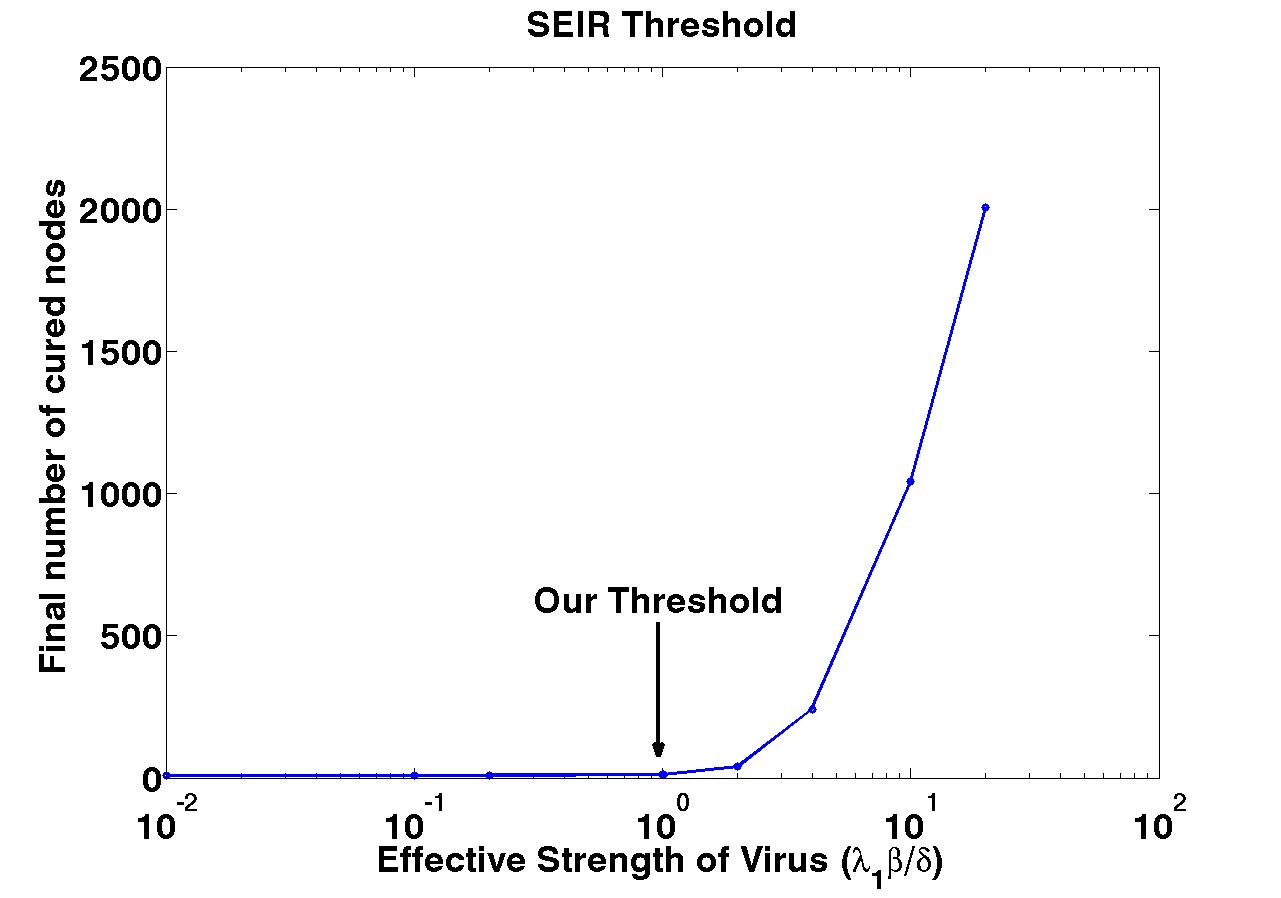} \\
	(A) SEIR Infective Fraction Time Plot (log-log) & (B) SEIR Footprint vs Strength (lin-log)
  \end{tabular}
\caption{\textbf{SEIR (all values averages over several runs): (A) Plot of Infective Fraction of Population vs Time (log-log). Note the qualitative difference in behavior \textit{under} and \textit{above} the threshold. Also notice the initial ``silent'' period for above threshold because of virus-incubation. (B) Plot of Final number of cured nodes (the \textit{footprint}) vs Effective Strength (lin-log). Note the \tps is exactly when the effective strength $s=1$.}}
\label{fig:infec-dynm-seir}
\end{figure}

\begin{figure}
\centering
    \begin{tabular}{cc}
	\includegraphics[width=3in]{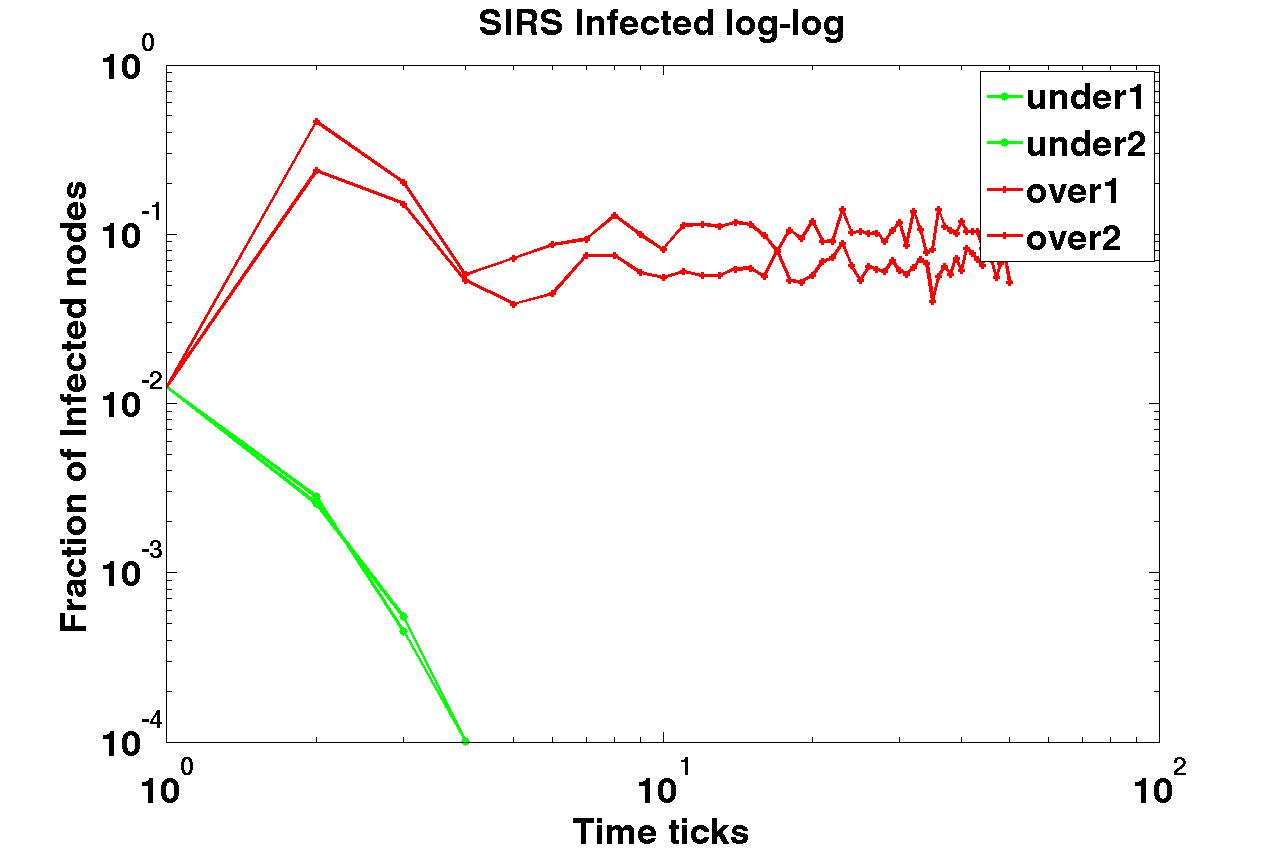} &
        \includegraphics[width=2.9in]{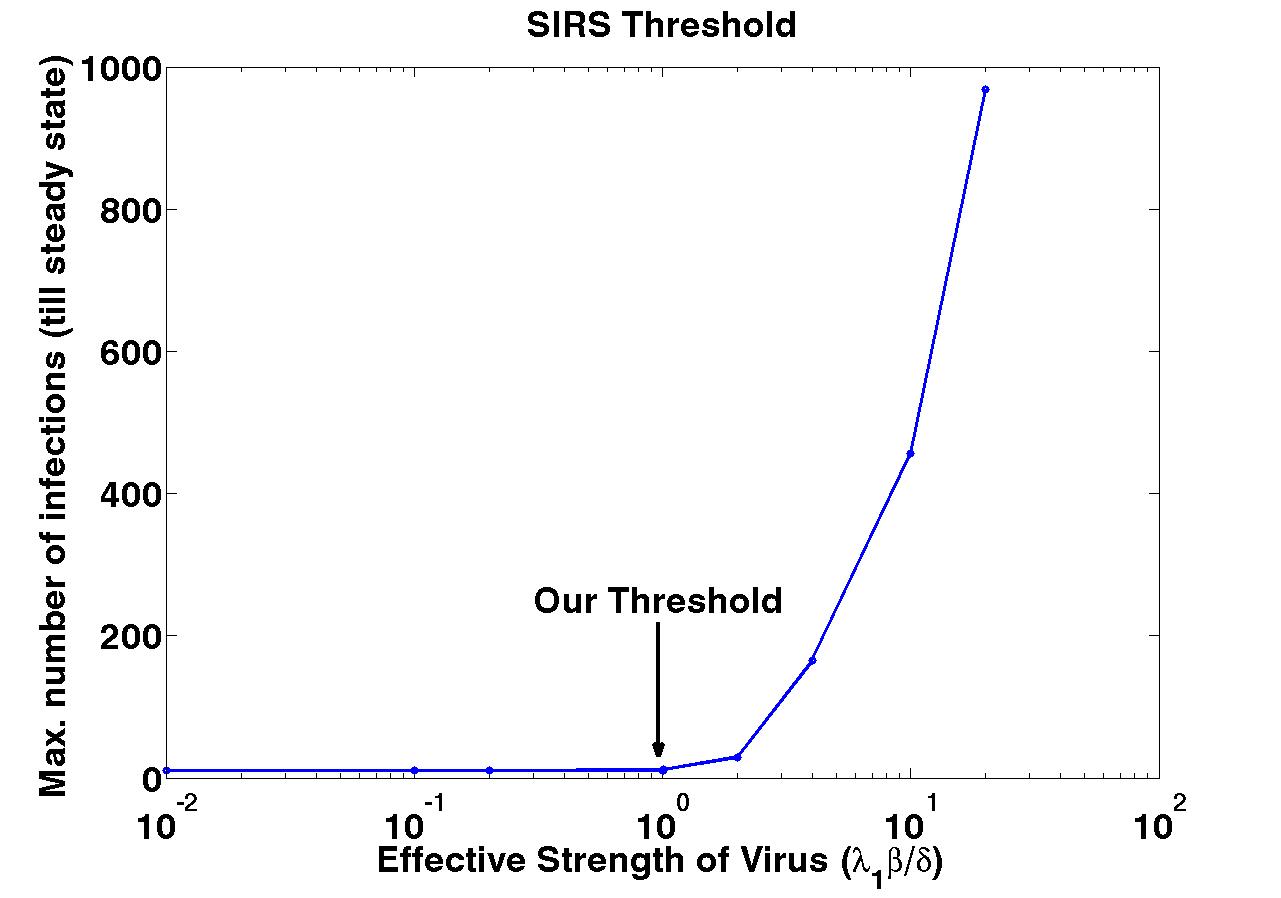} \\
	(A) SIRS Infective Fraction Time Plot (log-log) & (B) SIRS Max. Infections till steady \\
	&  state vs Strength (lin-log)
  \end{tabular}
\caption{\textbf{SIRS (all values averages over several runs): (A) Plot of Infective Fraction of Population vs Time (log-log). Note the qualitative difference in behavior \textit{under} and \textit{above} the threshold. (B) Plot of Max. number of infected nodes till steady state vs Effective Strength (lin-log). Note the \tps is exactly when the effective strength $s=1$.}}
\label{fig:infec-dynm-sirs}
\end{figure}

\begin{figure}
\centering
    \begin{tabular}{cc}
	\includegraphics[width=3in]{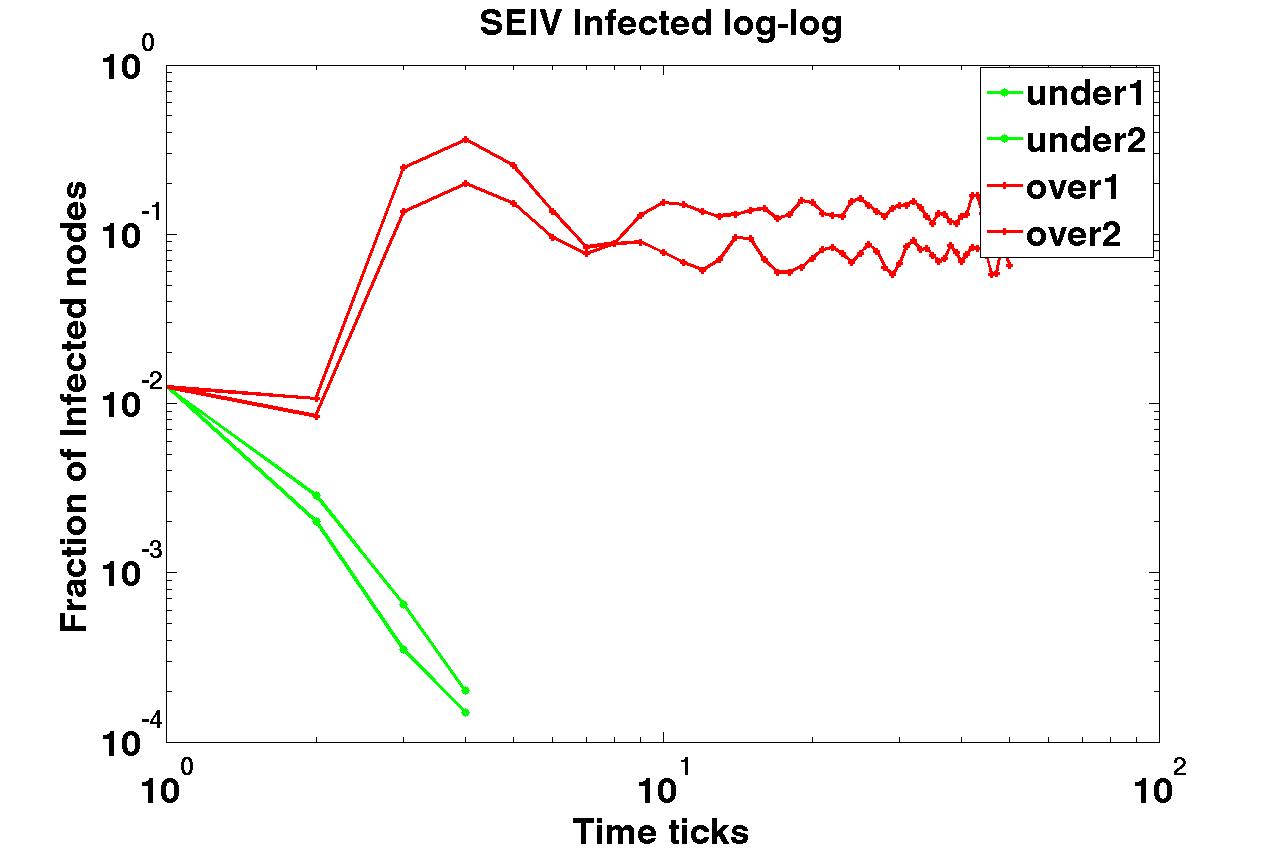} &
        \includegraphics[width=2.9in]{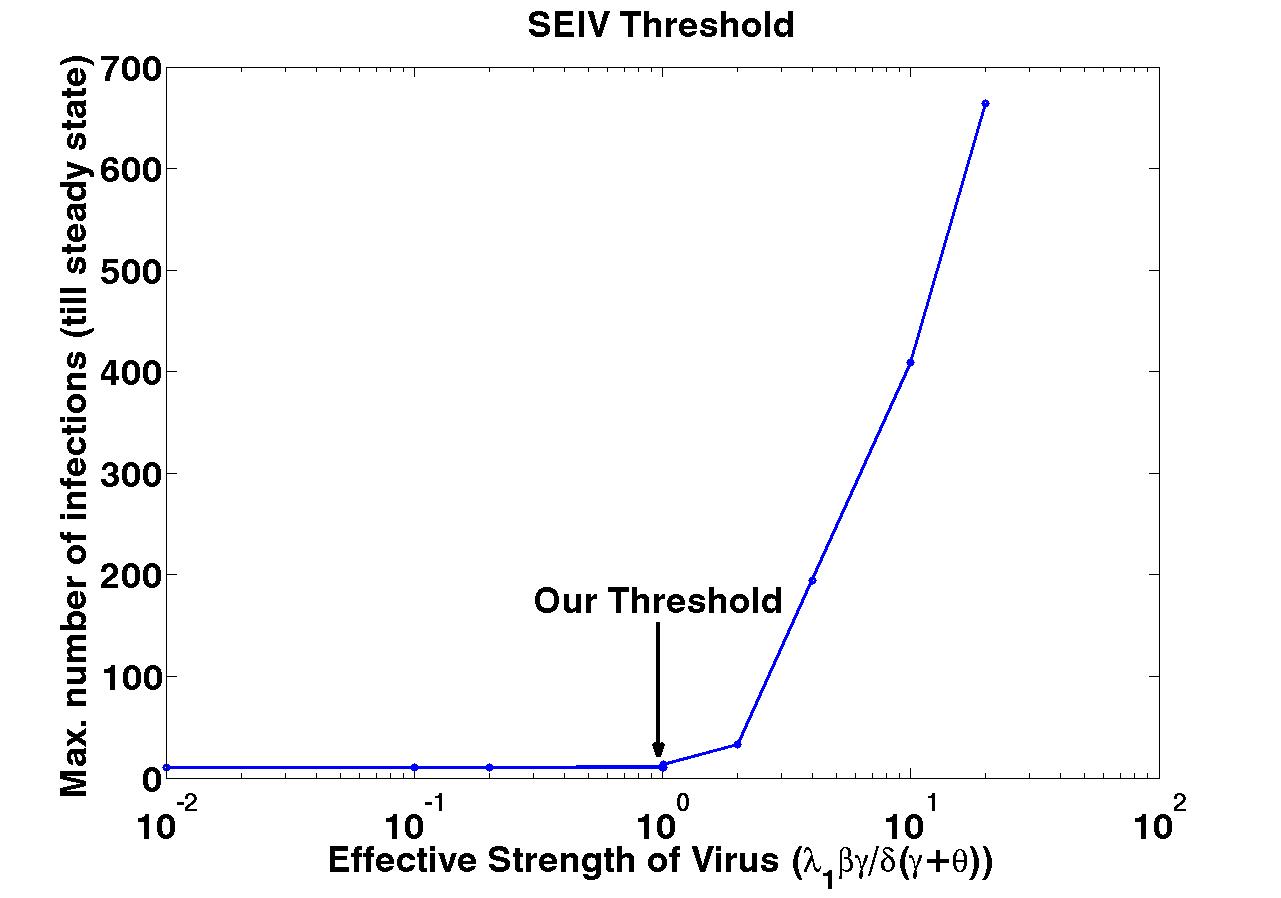} \\
	(A) SEIV Infective Fraction Time Plot (log-log) & (B) SEIV Max. Infections till steady \\
	&  state vs Strength (lin-log)
  \end{tabular}
\caption{\textbf{SEIV (all values averages over several runs): (A) Plot of Infective Fraction of Population vs Time (log-log). Note the qualitative difference in behavior \textit{under} and \textit{above} the threshold. Also notice the initial ``silent'' period for above threshold because of virus-incubation. (B) Plot of Max. number of infected nodes till steady state vs Effective Strength (lin-log). Note the \tps is exactly when the effective strength $s=1$.}}
\label{fig:infec-dynm-seiv}
\end{figure}
We conducted some computer simulation experiments on the Oregon AS router graph\footnote{This is a real network graph collected from the Oregon router views. It contains 15,420 links among 3,995 AS peers. More information can be found from \url{http://topology.eecs.umich.edu/data.html}.} to illustrate our \supermodel\ theorem. Figures~\ref{fig:infec-dynm-sir}, \ref{fig:infec-dynm-seir}, \ref{fig:infec-dynm-sirs} and \ref{fig:infec-dynm-seiv} give an overview of the simulations for the SIR, SEIR, SIRS and SEIV models. All values are average over several runs of the simulations. In short, as expected from the theorem, the difference in behavior above, below and at threshold can be distinctly seen for each of these models.  

Figures~\ref{fig:infec-dynm-sir}(A), \ref{fig:infec-dynm-seir}(A), \ref{fig:infec-dynm-sirs}(A) and \ref{fig:infec-dynm-seiv}(A) show a time-evolution plot of the fraction of infected nodes in the graph for different values of the effective strength of the virus specifically \textit{above threshold} (in red) and \textit{under threshold} (in green). Note the qualitative difference in the behavior of the system above and below the threshold. 

Figure~\ref{fig:infec-dynm-sir}(A) and Figure~\ref{fig:infec-dynm-seir}(A) deal with SIR and SEIR which don't have a steady state because of the recovered state. The number of susceptibles available in the graph decrease with each new infection and hence epidemics will disappear unlike in models like SIS and SIRS. Additionally as a result, above the threshold, the number of infections has an explosive phase and then they go down to zero while in the SIRS and SEIV models (see Figures~\ref{fig:infec-dynm-sirs}(A) and \ref{fig:infec-dynm-seiv}(A)) the number of infections reach a steady state value. Contrast this with the under threshold behavior, where in all the models the number of infections aggresively go down to zero. 

Also note the initial ``flat'' period in the time plots for above threshold for the models having the Exposed ($E$) state, SEIR and SEIV. This is due to the virus-incubation period because of which there is an initial delay in number of infected nodes. This then results in an initial ``silent'' period after which the epidemic takes-off. As there is no such incubation period in SIR and SIRS, their plots do not show such silent periods.

We also give a ``take-off'' plot for each model ((Figures~\ref{fig:infec-dynm-sir}(B), \ref{fig:infec-dynm-seir}(B), \ref{fig:infec-dynm-sirs}(B) and \ref{fig:infec-dynm-seiv}(B)). They show the final number of cured nodes (for SIR and SEIR) and max. number of infections till steady state (for SIRS and SEIV) vs the different strengths of the virus. Intuitively, these metrics measure the ``footprint'' of each infection. If the infection resulted in an epidemic then the footprint will be large. As our theorem predicted, the plots clearly illustrate that the \tps in all the cases is at the point when the effective strength $s=1$.

\subsection{$\eig$: Measure of Connectivity}
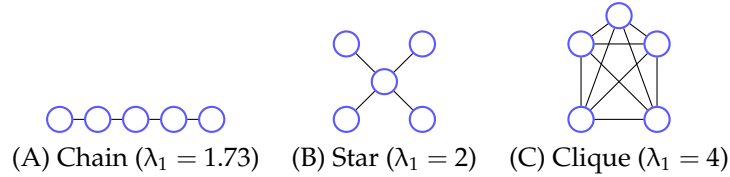
\begin{figure}[htb]
    \centering
    \begin{tabular}{ccc}
     \begin{tikzpicture}[scale=0.5]
      \tikzstyle{every node}=[draw=blue!65,circle,thick]
	\node at (-2,0) (n1) {};
	\node at (-1,0) (n2) {};
	\node at (0,0) (n3) {};
	\node at (1,0) (n4) {};
	\node at (2,0) (n5) {};
	\draw (n1) -- (n2);
	\draw (n2) -- (n3);	
	\draw (n3) -- (n4);
	\draw (n4) -- (n5);
     \end{tikzpicture}&

     \begin{tikzpicture}[scale=0.5]
      \tikzstyle{every node}=[draw=blue!65,circle,thick]
	\node at (0,0) (n1) {};
	\node at (-1,1) (n2) {};
	\node at (1,-1) (n3) {};
	\node at (1,1) (n4) {};
	\node at (-1,-1) (n5) {};
	\draw (n5) -- (n1);
	\draw (n2) -- (n1);	
	\draw (n3) -- (n1);
	\draw (n4) -- (n1);
     \end{tikzpicture}&

     \begin{tikzpicture}[scale=0.5]
      \tikzstyle{every node}=[draw=blue!65,circle,thick]
	\node at (0, 1.75) (n1) {};
	\node at (-1,1) (n2) {};
	\node at (1,-1) (n3) {};
	\node at (1,1) (n4) {};
	\node at (-1,-1) (n5) {};
	\draw (n5) -- (n1);\draw (n5) -- (n2);\draw (n5) -- (n3);\draw (n5) -- (n4);
	\draw (n2) -- (n1);\draw (n2) -- (n3);\draw (n2) -- (n4);	
	\draw (n3) -- (n1);\draw (n3) -- (n4);
	\draw (n4) -- (n1);
     \end{tikzpicture}\\

    (A) Chain ($\eig=1.73$) &(B) Star ($\eig=2$) & (C) Clique ($\eig=4$)\\  
    \end{tabular}
    \caption{\textbf{Changing connectivity and vulnerability of graphs with changing $\eig$. Our result says that a virus has the most effective strength in a clique, star and chain, in that order. Thus the Clique is the most vulnerable.}}
\label{fig:changinglambda}
\end{figure}
What does exactly the result mean w.r.t. the graph? 
Intuitively, $\eig$ (also known as the spectral radius) of a graph captures 
the connectivity of the graph. 
The preceding threshold results show 
that this is precisely what matters in the epidemic threshold. 
Hence, more connected the graph is, 
more vulnerable it is to an epidemic by a virus. 
For example, see Figure~\ref{fig:changinglambda}. 
It shows three graphs with the same number of nodes 
in increasing $\eig$ value. 
Note that although the star and chain 
(Figure~\ref{fig:changinglambda} (A) and (B)) 
have the same number of edges, 
the star has a higher $\eig$ value, 
thus boosting the effective strength $s$ 
of a virus making the graph more vulnerable to epidemics.
Also, recall that a $d$-regular graph (a homogeneous graph with all nodes of same degree $d$) has $\eig = d$. Hence, we can make the following observation: 
\begin{observation}[Impact of $\eig$]
Our threshold results suggest that an arbitrary graph behaves in the same way to a $\eig$-regular graph. In other words, $\eig$ captures the average neighborhood for a node in the graph. 
\end{observation}
Notice that the actual dynamics of the epidemic may not be captured by $\eig$ completely, but the \textit{threshold} is solely dependent on the $\eig$ (apart from the VPM). 

\subsection{Insensitivity to Virus Incubation}
\begin{figure}[htb]
\centering
  \includegraphics[width=4in]{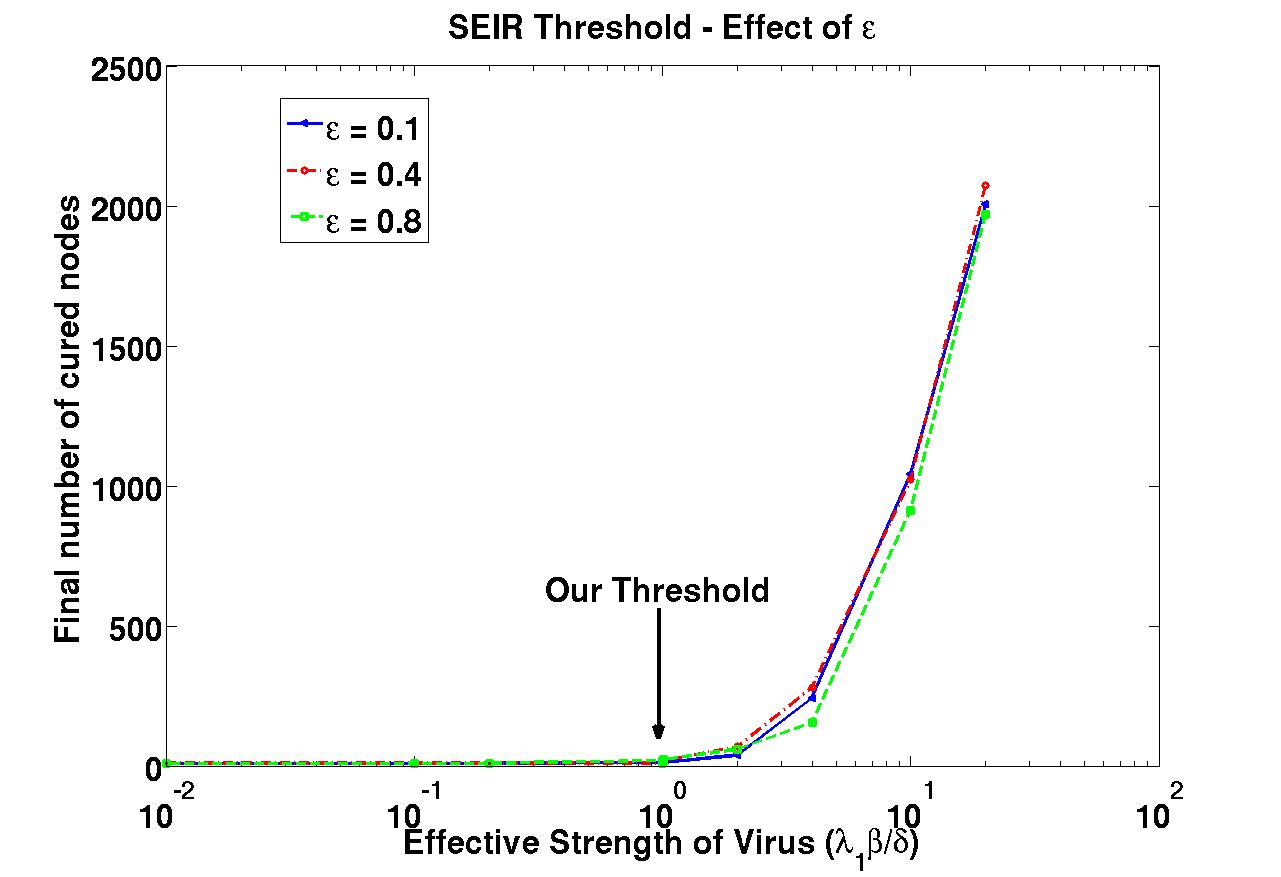} %
\caption{\textbf{Changing $\epsilon$ in the SEIR model on the AS graph (lin-log scale). Note that it \textit{does not} affect the threshold (the \tp\ is still at effective strength $\frac{\eig \beta}{\delta} = 1$). All values are averages over multiple runs.}}
\label{fig:seir-epsilon}
\end{figure}

For the SEIV model which is similar to SIV, except for the \exps\ state, we have the following:
\begin{observation}[SEIV]
The threshold here does not depend on the virus-maturation probability $\epsilon$ i.e.
the probability of transition from \textbf{E}xposed to \textbf{I}nfected. 
\end{observation}
This implies that the incubation period of the virus does not have an effect for purposes of the epidemic threshold. The parameter $\epsilon$, in effect, 
only delays/speeds-up the achievement of the threshold, 
not what the threshold itself is. As an example, this is also illustrated in Figure~\ref{fig:seir-epsilon} for the standard SEIR model~\cite{hethcote2000} (it is also one of the special cases of our SEIV generalization). 
It shows the final number of cured nodes (the `footprint') 
of computer simulations of the entire virus infection as a function of the effective strength 
of the virus ($\eig\beta/\delta$) on the AS graph. 
Note that the \tp\ is the \textit{same} 
for all the three colored curves 
which correspond to different values of $\epsilon$. 

\subsection{Confirming ``Prevention is better than cure''}

Clearly SIV is the most general model assuming no intermediate Exposed state and comprising three states (see Figure~\ref{fig:mh}). 
We know its effective strength (Table~\ref{tab:threshold}) is:
\[s = \eig \cdot \left(\frac{\beta\gamma}{\delta(\gamma+\theta)}\right)\]
By setting $\gamma$, $\theta$ etc. to zero we can get different models 
such as SIRS, SIR, SIS and so on. 
For example, when we set $\theta = 0$, the model reduces to SIRS, 
where the threshold we know is $\eig \beta / \delta$. This generalization gives us further insight:

\begin{observation}[Rate of loss of immunity]
Lowering the rate of loss of immunity i.e. having a smaller $\gamma$ (say due to better hygiene)   
decreases the effective strength $s$ 
(and makes it harder for the virus to cause an epidemic) 
only so long as there is a mechanism to give a node direct immunity i.e. having a non-zero $\theta$ 
(say by using a vaccine) \textit{before} an infection 
(in the Susceptible state) instead of \textit{after} 
(in the Recovered state). 
\end{observation}
Satisfyingly, this fits well with the old adage 
`Prevention is better than Cure'. This can also be vividly seen specifically in the case of SIRS ($\theta = 0$):
\begin{figure}[htbp]
\centering
 \includegraphics[width=4in]{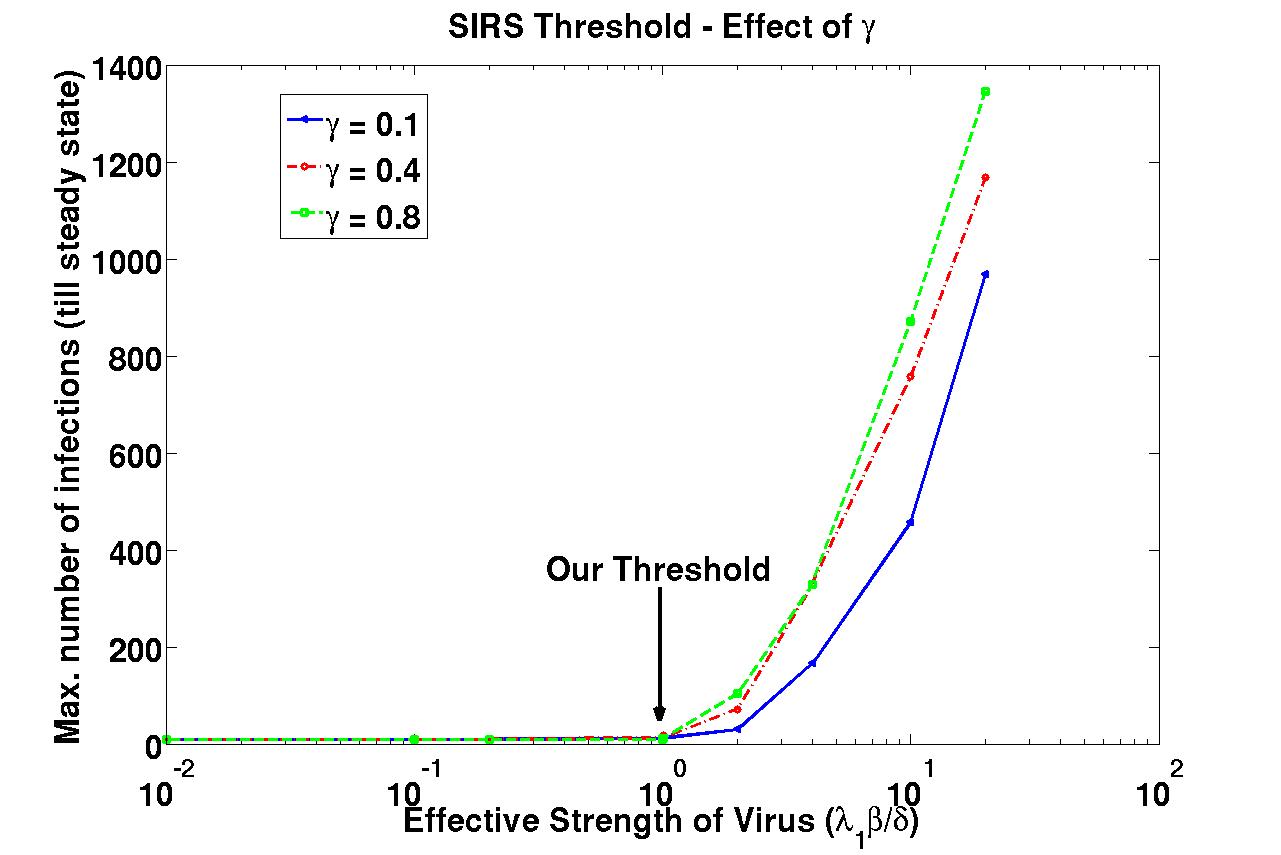}
\caption{\textbf{Changing $\gamma$ in the SIRS model on the AS graph (lin-log scale). Note that it \textit{does not} affect the threshold (the \tp\ is still at effective strength $\frac{\eig \beta}{\delta} = 1$). All values are averages over multiple runs.}}
\label{fig:sirs-gamma}
\end{figure}

\begin{observation}[SIRS]
The threshold in SIRS does not depend on $\gamma$, 
the immunity-loss rate, that is,
the probability of becoming \textbf{S}usceptible 
again from the \textbf{R}ecovered state. 
\end{observation}
Figure~\ref{fig:sirs-gamma} demonstrates this on the AS graph. 
Again, the \tp\ does not depend on $\gamma$, 
it is the same for all the three colored curves. 
This means that the rate of lapse or loss of immunity to a virus is immaterial to the threshold here.

\section{Conclusion}
\label{sec:conclusion}
In this article we provided two, orthogonal generalizations 
of earlier epidemic threshold  results. 
\begin{itemize*}
\item In the first direction,
    our result gives the threshold
    of the generalized \siiv\ model,
    which encompasses {\em any} 
    epidemic model in published 
    literature~(\cite{hethcote2000} \cite{networksbook} etc.). 
\item In the second direction, topology,
   we showed that for any, arbitrary, undirected contact-network,
   the effect of the topology can be captured solely by $\eig$,
   the first eigenvalue of the adjacency matrix. 
\end{itemize*}
We also demonstrated simulation results which illustrate our result. In addition, we discussed some important applications and implications of our result for policy makers, scientists etc. like: 
\begin{itemize*}
 \item Fast answers to ``what-if'' questions.
\item Guiding immunization policies.
\item Simplifying epidemiological simulations. 
\item Modeling Viral Marketing, Software patch distribution, Blog propagation etc.
\end{itemize*}

\section*{Acknowledgements}
This material is based upon work supported by
the National Science Foundation under Grants No. CNS-0721736 and CNS-0721889 and a Sprint gift.
Any opinions, findings, and conclusions or recommendations expressed in this
material are those of the authors and do not necessarily reflect
the views of the National Science Foundation, or other funding parties.

\bibliographystyle{abbrv}
\bibliography{BIB/paper}

\begin{thebibliography}{10}

\bibitem{andersonmay}
R.~M. Anderson and R.~M. May.
\newblock {\em Infectious Diseases of Humans}.
\newblock Oxford University Press, 1991.

\bibitem{barratbook}
A.~Barrat, M.~Barth\'elemy, and A.~Vespignani.
\newblock {\em Dynamical Processes on Complex Networks}.
\newblock Cambridge University Press, 2010.

\bibitem{bass69}
F.~M. Bass.
\newblock A new product growth for model consumer durables.
\newblock {\em Management Science}, 15(5):215--227, 1969.

\bibitem{deepay2008}
D.~Chakrabarti, Y.~Wang, C.~Wang, J.~Leskovec, and C.~Faloutsos.
\newblock Epidemic thresholds in real networks.
\newblock {\em ACM TISSEC}, 10(4), 2008.

\bibitem{networksbook}
D.~Easley and J.~Kleinberg.
\newblock {\em Networks, Crowds, and Markets: Reasoning About a Highly
  Connected World}.
\newblock Cambridge University Press, 2010.

\bibitem{ganesh2005}
A.~Ganesh, L.~Massoulie, and D.~Towsley.
\newblock The effect of network topology in spread of epidemics.
\newblock {\em IEEE INFOCOM}, 2005.

\bibitem{hayashi03}
Y.~Hayashi, M.~Minoura, and J.~Matsukubo.
\newblock Recoverable prevalence in growing scale-free networks and the
  effective immunization.
\newblock {\em arXiv:cond-mat/0305549 v2}, Aug. 6 2003.

\bibitem{hethcote2000}
H.~W. Hethcote.
\newblock The mathematics of infectious diseases.
\newblock {\em SIAM Review}, 42, 2000.

\bibitem{hethcote1984}
H.~W. Hethcote and J.~A. Yorke.
\newblock Gonorrhea transmission dynamics and control.
\newblock {\em Springer Lecture Notes in Biomathematics}, 46, 1984.

\bibitem{hirschsmale}
M.~W. Hirsch and S.~Smale.
\newblock {\em Differential Equations, Dynamical Systems and Linear Algebra}.
\newblock Academic Press, 1974.

\bibitem{matrixbook}
R.~A. Horn and C.~R. Johnson.
\newblock {\em Topics in Matrix Analysis}.
\newblock Cambridge University Press, 1991.

\bibitem{kephart1991}
J.~O. Kephart and S.~R. White.
\newblock Directed-graph epidemiological models of computer viruses.
\newblock {\em IEEE Computer Society Symposium on Research in Security and
  Privacy}, 1991.

\bibitem{kephart1993}
J.~O. Kephart and S.~R. White.
\newblock Measuring and modeling computer virus prevalence.
\newblock {\em IEEE Computer Society Symposium on Research in Security and
  Privacy}, 1993.

\bibitem{kleinberg07}
J.~Kleinberg.
\newblock The wireless epidemic.
\newblock {\em Nature, Vol. 449}, Sep 2007.

\bibitem{leskovec09}
J.~Leskovec, L.~Backstrom, and J.~Kleinberg.
\newblock Meme-tracking and the dynamics of the news cycle.
\newblock {\em ACM SIGKDD}, 2009.

\bibitem{mcculer2000}
C.~R. McCuler.
\newblock The many proofs and applications of perron's theorem.
\newblock {\em SIAM Review}, 42, 2000.

\bibitem{amspolyroots}
H.~W. Milnes.
\newblock Conditions that the zeros of a polynomial lie in the interval [-1, 1]
  when all zeros are real.
\newblock {\em The American Mathematical Monthly}, 70, No. 7, Aug. - Sept.
  1963.

\bibitem{vespignani2001}
R.~Pastor-Santorras and A.~Vespignani.
\newblock Epidemic spreading in scale-free networks.
\newblock {\em Physical Review Letters 86}, 14, 2001.

\bibitem{msr-worm}
M.~Vojnovic, V.~Gupta, T.~Karagiannis, and C.~Gkantsidis.
\newblock Sampling strategies for epidemic-style information dissemination.
\newblock {\em IEEE INFOCOM}, 2008.

\end{thebibliography}
\newpage
\appendix
\section*{Appendix}

\section{Notation}
\label{sec:notation}
Recall that we are dealing with the \siiv\ generalized model - it has two states $I_1$ and $I_2$ in the Infected class. To simplify notation, we refer to state $I_1$ as $E$ (the `infection entrance state') and $I_2$ as the $I$ state in the proofs. The state $E$ has a transmission probability of $\beta_1$ and the state $I$ has a transmission probability of $\beta_2$. The states $E$ and $I$ here should be thought as to mean \textit{general} infected states of our model and not in the sense of the specific $E$ and $I$ states in epidemic models like SEIR, SEIV etc. We also refer to the exogenous transitions as \textit{graph-based} and endogenous transitions as \textit{internal} interchangeably. Table~\ref{tab:mathnotation} gives some of the notation we will be using in our description of the proof.

\begin{table}[htb]
\caption{\textbf{Notation and Symbols used in proofs}} \label{tab:mathnotation}
\centering
\begin{small}\begin{tabular}{||l|p{4in}||}
\hline \hline
$m$ & total number of states in the model\\ \hline 
$q$ & total number of states in the Susceptible and Vigilant classes of the model; hence $m = q+2$ \\ \hline 
$w$ & total number of states in the Susceptible class of the model\\ \hline
$S_1, S_2, \ldots, S_w$ & general states in the Susceptible class \\ \hline
$E, I$ & general states in the Infected class \\ \hline
$\alpha_{KU}$ & probability (constant and given) of transition from state $K$ to state $U$ \\ \hline
$\beta_1$ & transmission probability for state $E$ \\ \hline
$\beta_2$ & transmission probability for state $I$ \\ \hline
$\zeta_{i,t}(E, I)$ & probability that a node $i$ does not receive any infections from $E$ and $I$ at time $t$ \\ \hline
$\vect{x}$ & the fixed point vector our NLDS corresponding to when no node is in any of the Infected class states \\ \hline
$p^*_{S_y}$ & (same for each node) probability of being present in the $S_y$ state at $\vect{x}$ \\ \hline
$\calJ$ & Jacobian matrix of the NLDS computed at  $\vect{x}$ \\ \hline
\end{tabular}\end{small}
\end{table}

\section{System Equations}
\label{sec:syseqn}
We can develop the system equations i.e. explicitly specify the non-linear function $g$ for the NLDS based on the transition diagram of the model. 
As stated earlier in Section~\ref{sec:genmodel} we assume that infections are received only from infected neighbors i.e. those in states $E$ and $I$ the Infected class of states. Firstly, let's calculate the probability that a node $i$ does not receive any infections in the next time step (call it $\zeta_{i,t}(E, I)$, $E, I$ denotes that an infection is passed only from a neighbor in the $E$ or $I$ states). No infections are transmitted if:
\begin{itemize*}
\item Either a neighbor is not any of the infected states $E$ and $I$
\item Or it is in state $E$ and the transmission fails with probability $1 - \beta_1$ 
\item Or it is in state $I$ and the transmission fails with probability $1 - \beta_2$
\end{itemize*}
Since we assume infinitesimally small time steps ($\Delta t \rightarrow 0$), multiple events can be ignored for first-order effects in the time step. Also, assuming the neighbors are \textit{independent}, we get:
\begin{eqnarray}
\zeta_{i,t}(E, I)& = &\prod_{j \in {\cal NE}(i)} \left(P_{E,j,t} (1 - \beta_1) + P_{I,j,t} (1 - \beta_2) + (1 - P_{E,j,t} - P_{I,j,t}) \right) \nonumber \\
& =  &\prod_{j\in \{1..N\}} \left(1 - \matA_{i,j}(\beta_1 P_{E,j,t} + \beta_2 P_{I,j,t}) \right) \label{eq:zeta}
\end{eqnarray}
where ${\cal NE}(i)$ is the set of neighbors of node $i$ in the graph. 

Also, the sum of probabilities of being in all the possible states for each node $i$ should equal $1$. Hence,
\begin{equation}
 \forall_{i,t}~~~ \sum_{K} P_{K,i,t} = 1
\end{equation}

We can now write down the system equations as follows. A node $i$ will be in any particular state $S_y$ of the Susceptible class at time $t+1$ if:
\begin{itemize*}
 \item Either it was in $S_y$ at time $t$ and stayed in state $S_y$ i.e. it did not receive any infections from its neighbors \textit{and} it did not change state internally from $S_y$ to any other state  
\item Or it was in some other state $U$ and changed state internally from $U$ to $S_y$
\end{itemize*}

Hence, the probability of node $i$ being in $S_y$ where $S_y$ is any state in the Susceptible class at time $t+1$ is:
\begin{equation}
\label{eq:susstate}
\forall y = {1, 2, \ldots, w}~~~P_{S_y,i,{t+1}} = \sum_{K\neq {S_y}}\alpha_{KS_y} P_{K,i,t}  + P_{S_y, i, t} \left(\zeta_{i,t}(E, I) - \sum_{K\neq {E,S_y,I}} \alpha_{S_yK}\right)
\end{equation}

Similarly, for the $E$ state:
\begin{equation}
\label{eq:expstate}
P_{E,i,{t+1}} = \sum_{K\neq {S_1, S_2, \ldots, S_w}}\alpha_{KE} P_{K,i,t} + \sum_{y = 1}^{w} P_{S_y, i, t} \left(1 - \zeta_{i,t}(E, I)\right)
\end{equation}

and for any other state $U \neq \{S_1, S_2, \ldots, S_w, E\}$:
\begin{equation}
\label{eq:mstate}
P_{U,i,{t+1}} = \sum_{K}\alpha_{KU} P_{K,i,t} 
\end{equation}

As discussed earlier (Equation~\ref{eq:probvec}), we can now define a probability vector $\vect{P}_t$ by ``stacking'' all these probabilities which will completely describe the system at any time $t$ and evolve according to the above equations. Note that the above equations are non-linear and naturally define the function $g$ for the NLDS $\vect{P}_{t+1} = g(\vect{P}_t)$. 

We have the following theorem about NLDS stability at a fixed point:

\begin{theorem}[Asymptotic Stability, e.g. see~\cite{hirschsmale}]
\label{theorem:asymstable}
The system given by $\vect{P}_{t+1} = g(\vect{P}_t)$ is asymptotically stable at an equilibrium point $\vect{P} = \vect{x}$, 
if the eigenvalues of $\calJ = \bigtriangledown g(\vect{x})$ 
are less than 1 in absolute value, 
where, 
\[ \calJ_{i,j}  = \left[\bigtriangledown g(\vect{x}) \right]_{i,j} = \frac{\partial g_i}{\partial g_j}|_{\vect{P}=\vect{x}}\] 
\end{theorem}

Hence, next we compute the fixed point we are interested in and the Jacobian of our NLDS at that point.

\section{Fixed point}
\label{sec:fixedpt}
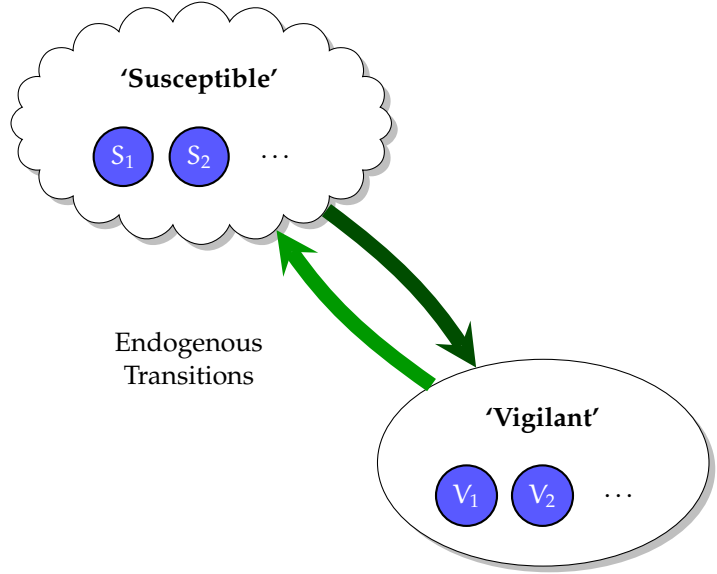
\begin{figure}[htb]
\begin{centering}
\begin{tikzpicture}[remember picture]
\node[cloud, draw, cloud puffs=20, aspect=1.75, cloud puff arc=140,fill=white,drop shadow] at (-4.5, 0) (sus)
	{\begin{tikzpicture}[place/.style={fill=blue!65,draw,circle,thick,text=white, text badly centered}]
	\node[place] at (-1, 0) (S1) {$S_1$}; 
	\node[place] at (0, 0) (S2) {$S_2$}; 
	\node at (1, 0) (S3) {$\ldots$}; 
	\node[rectangle]  at (0,1) {\textbf{`Susceptible'}};
	\end{tikzpicture}};
\node[ellipse, draw,fill=white,drop shadow] at (0, -4.5) (vigi)
	{\begin{tikzpicture}[place/.style={fill=blue!65,draw,circle,thick,text=white, text badly centered}]
	\node[place] at (-1, 0) (V1) {$V_1$}; 
	\node[place] at (0, 0) (V2) {$V_2$}; 
	\node at (1, 0) (V3) {$\ldots$}; 
	\node[rectangle]  at (0,1) {\textbf{`Vigilant'}};
	\end{tikzpicture}};
\end{tikzpicture}
\begin{tikzpicture}[remember picture, overlay]
\begin{scope}[>=stealth]
\draw[green!60!black, line width=2mm, ->] (vigi) to [bend left=10] (sus)
node[auto,text=black,midway,text width=3.7cm,text centered] {Endogenous Transitions};
\draw[green!30!black, line width=2mm, ->] (sus) to [bend left=10] (vigi);
\end{scope}
\end{tikzpicture}
\caption{\textbf{State Diagram for any node in the graph at the fixed point when \textit{no} node is present in a state in the Infected class. Only cross-class edges are shown. Note that it is now a simple Markov chain with a unique steady state probability.}}
\label{fig:statediafixedpoint} 
\end{centering}
\end{figure}

We are interested in the stability of the equilibrium point (i.e. where $\vect{P}_{t+1} = \vect{P}_t (= \vect{x})$) of the NLDS which corresponds to when no one is infected. Only the transition from the Susceptible class states towards the Infected class states are graph-based (and can happen only when at least one of the nodes is in any of the Infected states), so the state-diagram for each node will be a simple Markov chain (call it $MC_{SV}$) consisting of the Susceptible and Vigilant states (see Figure~\ref{fig:statediafixedpoint}). Note now there are no graph-based effects, hence each node is independent of others and will converge to steady state probabilities corresponding to the Markov chain. The steady state vector $\pi^*$ (size $q \times 1$, where $q$ is the number of states in the Susceptible and Vigilant classes) which will be the same for each node can be computed from the following equations from standard Markov chain analysis:
\begin{eqnarray}
\pi^{*T}~\mat{\mathrm{Tran}}_{MC_{SV}} = \pi^* \mathrm{~~\&~~} \sum_{i=1}^q {\pi^*_i} = 1
\end{eqnarray}

Hence $\pi^*$ is a probability vector and is the left eigenvector corresponding to eigenvalue 1 of the stochastic matrix $\mat{\mathrm{Tran}}_{MC_{SV}}$ of the Markov chain $MC_{SV}$. The full ($m \times 1$) probability vector $\vect{p^*}$ for each node at this steady state will have the entries in $\pi^*$ for states in the Susceptible and Vigilant classes and 0 for all states in the Infected class. The fixed point of the global original NLDS $\vect{x}$ can be finally represented as:
\begin{equation}
 \vect{x} = \left[ \begin{matrix} 
                    \vect{p^*} \\
		    \vect{p^*}\\
		     \vdots \\
		     \vect{p^*}
                    \end{matrix} \right]
\end{equation}
where $\vect{p^*}$ is repeated $N$ times (once for each node in the graph). Let $p^*_{S_y}$ be the steady state probability value in the vector $\vect{p^*}$ corresponding to the $S_y$ state. In other words, each node will have a probability of $p^*_S$ of being present in the $S_y$  state at the fixed point. Also define, 
\begin{equation}
p^*_S = \sum_{y = 1}^{w} p^*_{S_y}
\end{equation} 
i.e. $p^*_S$ is the total probability of each node at the fixed point of being present in any of the states of the Susceptible class. 
\section{The Jacobian}
\label{sec:jacobian}
We know from Theorem~\ref{theorem:asymstable} that $\vect{x}$ is stable if the eigenvalues of $\calJ = \bigtriangledown g(\vect{x})$ are less than 1 in absolute value. From the definition of $\calJ$ we can see that it is a $m\cdot N \times m\cdot N$ matrix with $m$ (for each state) square blocks of size $N \times N$ each (corresponding to every node in the graph). We can calculate $\calJ$ to be (states have been mentioned on the top and side for ease of exposition and $\eye$ is the identity matrix of size $N \times N$):

\begin{center}
\renewcommand{\arraystretch}{1.5}
\begin{tabular}{c|c|c|c|c|c|}
& $S_y$ & $K$ & $\hdots$ & \exps\ & \infd\ \\  
\hline 
$S_y$ & $(1- \sum_{K \neq {S_y,E}} \alpha_{S_yK})\eye$ & $\alpha_{KS_y}\eye$ & $\hdots$ & $\alpha_{ES_y}\eye - p^*_{S_y}\beta_1 \matA$& $\alpha_{IS_y}\eye - p^*_S\beta_2 \matA$\\
\hline
$\vdots$&  &  &$\ddots$  &  &\\ 
\hline
$U$ & $\alpha_{S_yU}\eye$ & $\alpha_{KU}\eye$ & $\hdots$ & $\alpha_{EU}\eye$ & $\alpha_{IU}\eye$   \\
\hline
$\vdots$&  &  &$\ddots$  &  &\\ 
\hline
\exps\ & $\alpha_{S_yE}\eye$ & $\alpha_{KE}\eye$ & $\hdots$ & $\alpha_{EE}\eye + p^*_S\beta_1 \matA$ & $\alpha_{IE}\eye + p^*_S\beta_2 \matA$ \\ 
\hline
\infd\ & $\alpha_{S_yI}\eye$ & $\alpha_{KI}\eye$ & $\hdots$ & $\alpha_{EI}\eye$ & $\alpha_{II}\eye$   \\
\hline
\end{tabular}
\end{center}
where $K$ is any state $\neq \{E, I\}$ and $U$ is any state $\neq \{S_1, S_2, \ldots, S_w, E, I\}$. 

Recall the properties we are assuming for the epidemic models discussed in Section~\ref{sec:genmodel} (also see Figure~\ref{fig:statedia}). Crucially, they imply $\forall_{K \neq {E,I}}~\alpha_{KE} = 0$ and $\forall_{K \neq {E,I}}~\alpha_{KI} = 0$. Hence $\calJ$ reduces to: 
\begin{center}
\renewcommand{\arraystretch}{1.5}
\begin{tabular}{c|c|c|c|c|c|}
& $S_y$ & $K$ & $\hdots$ & \exps\ & \infd\ \\  
\hline 
$S_y$ & $(1- \sum_{K \neq {S_y,E}} \alpha_{S_yK})\eye$ & $\alpha_{KS_y}\eye$ & $\hdots$ & $\alpha_{ES_y}\eye - p^*_{S_y}\beta_1 \matA$& $\alpha_{IS_y}\eye - p^*_S\beta_2 \matA$\\
\hline
$\vdots$&  &  &$\ddots$  &  &\\ 
\hline
$U$ & $\alpha_{S_yU}\eye$ & $\alpha_{KU}\eye$ & $\hdots$ & $\alpha_{EU}\eye$ & $\alpha_{IU}\eye$   \\
\hline
$\vdots$&  &  &$\ddots$  &  &\\ 
\hline
\exps\ & $\matz_{N, N}$ & $\matz_{N, N}$ & $\hdots$ & $\alpha_{EE}\eye + p^*_S\beta_1 \matA$ & $\alpha_{IE}\eye + p^*_S\beta_2 \matA$ \\ 
\hline
\infd\ & $\matz_{N, N}$ & $\matz_{N, N}$ & $\hdots$ & $\alpha_{EI}\eye$ & $\alpha_{II}\eye$   \\
\hline
\end{tabular}
\end{center}
where $\matz_{N,N}$ is a $N \times N$ matrix with all zeros.

\section{Eigenvalues of the Jacobian}
\label{sec:eigenvalues}
Note that $\calJ$ is very structured and can be written as:
 \begin{equation}
\label{eq:jblock}
 \calJ = \left[ \begin{matrix} 
		\matB_1 & \matB_2 \\ \matz_{2N,(m-2)N} & \matB_3
                \end{matrix} \right]
\end{equation}
\noindent
where $\matB_1$, $\matB_2$ and $\matB_3$ are matrices of size $(m-2)N \times (m-2)N$, $(m-2)N \times 2N$ and $2N \times 2N$ respectively. $\matB_3$ corresponds to the \exps\ and \infd\ rows and columns of $\calJ$ i.e.:
\begin{equation}
 \matB_3 = \left[ \begin{matrix} 
		 \alpha_{EE}\eye + p^*_S\beta_1 \matA  & \alpha_{IE}\eye + p^*_S\beta_2 \matA \\ 
                 \alpha_{EI}\eye & \alpha_{II}\eye
                \end{matrix} \right]
\end{equation}
\noindent
$\matB_1$ and $\matB_2$ are defined similarly. Consider any eigenvector $\vect{v}$ (size $mN \times 1$) and corresponding eigenvalue $\lambdaJ$ of $\calJ$. We can write $\vect{v}$ as being composed of vector $\vect{v}_1$ of size $(m-2)N \times 1$ and vector $\vect{v}_2$ of size $2 N \times 1$ i.e:
\begin{equation}
\label{eq:jeigenvec}
 \vect{v} = \left[ \begin{matrix}
                    \vect{v}_1\\
                     \vect{v}_2\\  
		   \end{matrix} 
            \right]
\end{equation}
Also $\vect{x}$ and $\lambdaJ$ satisfy the eigenvalue equation: 
\begin{equation}
\calJ \vect{v} = \lambdaJ \vect{v}
\end{equation}
\noindent
Substituting from Equations~\ref{eq:jblock} and~\ref{eq:jeigenvec} we get:
\begin{equation}
\label{eq:eigen}
 \left[ \begin{matrix} 
		\matB_1 & \matB_2 \\ \matz & \matB_3
                \end{matrix} 
\right] \left[ \begin{matrix}
                    \vect{v}_1\\
                     \vect{v}_2\\  
                   \end{matrix} 
            \right]   = \lambdaJ \left[ \begin{matrix}
                    \vect{v}_1\\
                     \vect{v}_2\\  
                   \end{matrix} 
            \right]
\end{equation}
\noindent
Equation~\ref{eq:eigen} implies the following the two relations:
\begin{eqnarray}
 \matB_1 \vect{v}_1 + \matB_2 \vect{v}_2 & = & \lambdaJ \vect{v}_1 \label{eq:eigenB1B2} \\ 
  \matB_3 \vect{v}_2 &= & \lambdaJ \vect{v}_2 \label{eq:eigenB3}
\end{eqnarray}

\noindent
From Equation~\ref{eq:eigenB3} we can infer that precisely \textit{one} of the following holds:
\begin{enumerate*}
 \item $\vect{v}_2 = \vect{0}$
\item  $\vect{v}_2$ is the eigenvector of $\matB_3$ (and consequently $\lambdaJ$ is the matching eigenvalue of $\matB_3$)
\end{enumerate*}
\noindent
If $\vect{v}_2 = \vect{0}$, Equation~\ref{eq:eigenB1B2} reduces to \[\matB_1\vect{v}_1 = \lambdaJ \vect{v}_1\] wherein again, either $\vect{v}_1 = \vect{0}$ or $\lambdaJ$ is an eigenvalue of $\matB_1$. The condition $\vect{v}_1 = \vect{0}$ is not meaningful as then $\vect{v} = \vect{0}$ ($\vect{v}$ is an eigenvector of $\calJ$ implies $\vect{v}$ is non-zero). Therefore the eigenvalues of $\calJ$ are given by the eigenvalues of $\matB_1$ (with $\vect{v}_2 = \vect{0}$) and the eigenvalues of $\matB_3$.

\subsection{Eigenvalues of $\matB_1$}
From the expression for $\calJ$ derived in Section~\ref{sec:jacobian}, note that:
\begin{equation}
\label{eq:kron}
\matB_1 = \mat{T} \otimes \eye
\end{equation}
\noindent
where $\otimes$ is the Kronecker product of two matrices and
\begin{equation}
\label{eq:t}
\mat{T} = \left[ \begin{matrix} 
		(1- \sum_{K \neq {S_y,E}} \alpha_{S_yK}) & \alpha_{KS_y} & \hdots \\
		\vdots & \vdots & \vdots \\
		\alpha_{S_yU}& \alpha_{KU}& \hdots \\
		\vdots & \ddots & \vdots 
                \end{matrix} 
\right] 
\end{equation}
We know from matrix algebra~\cite{matrixbook} that if $\mat{C} = \mat{D} \otimes \mat{E}$ then $\mat{C}_\lambda = \mat{D}_\lambda \otimes \mat{E}_\lambda$, where $\mat{C}_\lambda$  denotes a  diagonal matrix with eigenvalues of the matrix $\mat{C}$ on the diagonal. But $\eye_\lambda = \eye$, hence the eigenvalues of $\matB_1$ are the same as the eigenvalues of $\mat{T}$ (although with repetition). In other words, eigenvalues of $\mat{T}$ are eigenvalues of $\calJ$ as well.

\subsection{Eigenvalues of $\matB_3$}
Let $\vect{u} = \left[ \begin{matrix} \vect{u}_1 \\ \vect{u}_2 \end{matrix} \right]$ be a corresponding eigenvector of $\matB_3$ ($\vect{u}_1$ and $\vect{u}_2$ are of size $N \times 1$ each and as the eigenvalues of $\matB_3$ are also eigenvalues of $\calJ$, we use $\lambdaJ$ for an eigenvalue of $\matB_3$). Hence, the standard eigenvalue relation $\matB_3 \vect{u} = \lambdaJ \vect{u}$  requires the following equations to be satisfied:
\begin{eqnarray}
(\alpha_{EE}\eye + p^*_S\beta_1 \matA) \vect{u}_1 + (\alpha_{IE}\eye + p^*_S\beta_2 \matA) \vect{u}_2 & = & \lambdaJ \vect{u}_1 \label{eq:b1first} \\ 
                 \alpha_{EI} \vect{u}_1 + \alpha_{II} \vect{u}_2 & = & \lambdaJ \vect{u}_2 \label{eq:b1sec}
\end{eqnarray}
\noindent
Using Equation~\ref{eq:b1sec}, we can compute $\vect{u}_1$ in terms of $\vect{u}_2$ as:
\begin{equation}
 \vect{u}_1 = \left( \frac{\lambdaJ - \alpha_{II}}{\alpha_{EI}} \right) \vect{u}_2
\end{equation}
\noindent
Substituting it back into Equation~\ref{eq:b1first} we get:
\begin{eqnarray}
 \left( (\alpha_{EE}\eye + p^*_S\beta_1 \matA) \left(\frac{\lambdaJ - \alpha_{II}}{\alpha_{EI}} \right) + \alpha_{IE}\eye + p^*_S\beta_2 \matA \right) \vect{u}_2 &=& \lambdaJ \left( \frac{\lambdaJ - \alpha_{II}}{\alpha_{EI}} \right) \vect{u}_2 \nonumber \\
\Rightarrow \left( \alpha_{EE} (\lambdaJ - \alpha_{II})\eye + \alpha_{IE} \alpha_{EI}\eye + (p^*_S\beta_1 (\lambdaJ - \alpha_{II}) + p^*_S\beta_2\alpha_{EI}) \matA \right) \vect{u}_2 & = & \lambdaJ (\lambdaJ - \alpha_{II}) \vect{u}_2 \nonumber
\end{eqnarray} 
which finally gives,
\begin{equation}
\matA \vect{u}_2  =  \left(\frac{\lambdaJ^2 - (\alpha_{II} + \alpha_{EE})\lambdaJ + \alpha_{II}\alpha_{EE} - \alpha_{IE}\alpha_{EI}}{p^*_S\beta_1 (\lambdaJ - \alpha_{II}) + p^*_S\beta_2\alpha_{EI}}\right) \vect{u}_2 \label{eq:finaleigen}
\end{equation}

Again, Equation~\ref{eq:finaleigen} tells us that either $\vect{u}_2 = \vect{0}$ or it is an eigenvector for $\matA$. But $\vect{u}_2 = \vect{0} \Rightarrow \vect{u}_1 = \vect{0} \Rightarrow \vect{u} = \vect{0}$  which is not possible. Thus Equation~\ref{eq:finaleigen} is an eigenvalue equation for the adjacency matrix $\matA$ and we are looking for solutions $\lambdaJ$ and $\vect{u}_2$ such that they satisfy it. Hence, 
\begin{equation*}
\lambdaA = \frac{\lambdaJ^2 - (\alpha_{II} + \alpha_{EE})\lambdaJ + \alpha_{II}\alpha_{EE} - \alpha_{IE}\alpha_{EI}}{p^*_S\beta_1 (\lambdaJ - \alpha_{II}) + p^*_S\beta_2\alpha_{EI}} \nonumber 
\end{equation*}
\noindent
where $\lambdaA$ is an eigenvalue of $\matA$. This finally gives
\begin{equation}
\lambdaJ^2 - \lambdaJ(\alpha_{EE} + \alpha_{II} + p^*_S\beta_1\lambdaA) + (\alpha_{II} \alpha_{EE} - \alpha_{IE}\alpha_{EI} +  p^*_S\lambdaA (\beta_1\alpha_{II}-\beta_2\alpha_{EI})) = 0 \label{eq:quad}
\end{equation}
\noindent
Thus we have a different quadratic equation (\qe) for each eigenvalue $\lambdaA$ of $\matA$. Each \qe\ gives us two eigenvalues (possibly repeated) of $\calJ$.

So, finally, we can conclude the following lemma:
\begin{lemma}[Eigenvalues of $\calJ$]
\label{lem:eigJ}
Eigenvalues of $\calJ$ are given by the eigenvalues of $\mat{T}$ (Equations~\ref{eq:kron} and~\ref{eq:t}) and the roots of the \qes\ given by Equation~\ref{eq:quad} for each eigenvalue $\lambdaA$ of $\matA$.  
\end{lemma}

\section{Stability}
\label{sec:stability}
We require that all the eigenvalues of $\calJ$ to be less than 1 in absolute value (according to Theorem~\ref{theorem:asymstable}). From Lemma~\ref{lem:eigJ}, we have two cases to handle in enforcing this:
\begin{enumerate*}
 \item[\textbf{(C1)}] All the eigenvalues of $\mat{T}$ should be less than 1 in absolute value
\item[\textbf{(C2)}] All the roots of the \qes\ given by Equation~\ref{eq:quad} for each eigenvalue $\lambdaA$ of $\matA$ should be less than 1 in absolute value
\end{enumerate*}
 
\subsection{Case C1}
Note that this case depends \textit{only} on the model as the matrix $\mat{T}$ is \textit{independent} of the adjacency matrix $\matA$. But $\mat{T}$ is a stochastic matrix i.e. all the column sums are equal to 1 - consequently all its eigenvalues are less than 1 in absolute value. 
\begin{lemma}[Stability C1]\label{lem:eigT}
All eigenvalues of the matrix $\mat{T}$ (given by Equation~\ref{eq:t}) are less than 1 in absolute value. 
\end{lemma}

\subsection{Case C2}
As \textbf{C1} is always true, we need to only ensure case~\textbf{C2}. We can prove here the following:

\begin{lemma}[Stability C2]\label{lem:eigquad}
All the roots of the \qes\ given by Equation~\ref{eq:quad} for each eigenvalue $\lambdaA$ of $\matA$ are less than 1 in absolute value if:
\[ \eig p^*_S \left(\frac{\beta_1 (1- \alpha_{II}) + \beta_2 \alpha_{EI}}{(1-\alpha_{II})(1- \alpha_{EE}) - \alpha_{IE}\alpha_{EI}}\right)  < 1 \]
\end{lemma}

\paragraph{Proof}
Let $r_1$ and $r_2$ be the roots of Equation~\ref{eq:quad} ($r_1$ and $r_2$ can be real or complex depending on $\lambdaA$). Then we want
\begin{equation*}
 |r_1| < 1 \mathrm{~and~} |r_2| < 1   
\end{equation*}

\paragraph{$r_1$ and $r_2$ are real}
As the roots are real, $\lambdaA$ is such that the discriminant $\calD$ of the quadratic equation is greater than zero. In this situation: 
\begin{equation}
|r_1| < 1 \mathrm{~and~} |r_2| < 1 \Rightarrow  r_1 \in (-1, 1)  \mathrm{~and~} r_2 \in (-1, 1) 
\end{equation}

From the theory of quadratic equations, it is well known (see e.g.~\cite{amspolyroots}) that for real roots $x_1$ and $x_2$ of a \qe\ $f(x) = a x^2 + b x + c$ (with $a > 0$) to lie in the interval $(-1, 1)$ the following conditions must be true:
\begin{align*}
a - c > 0, \\
a - b + c > 0,           \\
a + b + c > 0.             
\end{align*}
\noindent
Intuitively, the first condition forces the product of the roots to be less than 1 while the last two conditions state that value of $f(x)$ at $-1$ and $1$ should not be ``too small''. In our case, these then translate into:
\begin{subequations}
 \begin{align}
\alpha_{II} \alpha_{EE} - \alpha_{IE}\alpha_{EI} +  p^*_S\lambdaA (\beta_1\alpha_{II}-\beta_2\alpha_{EI}) < 1 \label{eq:qec1}   \\
1 + \alpha_{EE} + \alpha_{II} + p^*_S\beta_1\lambdaA + \alpha_{II} \alpha_{EE} - \alpha_{IE}\alpha_{EI} +  p^*_S\lambdaA (\beta_1\alpha_{II}-\beta_2\alpha_{EI}) > 0  \label{eq:qec2}  \\
1 - \alpha_{EE} - \alpha_{II} - p^*_S\beta_1\lambdaA + \alpha_{II} \alpha_{EE} - \alpha_{IE}\alpha_{EI} +  p^*_S\lambdaA (\beta_1\alpha_{II}-\beta_2\alpha_{EI}) > 0 \label{eq:qec3}
\end{align}
\end{subequations}
 
Equations~\ref{eq:qec2} and~\ref{eq:qec3} can be written as:
\begin{subequations}
 \begin{align}
  \lambdaA p^*_S \left(\frac{-\beta_1 (1+ \alpha_{II}) + \beta_2 \alpha_{EI}}{(1+\alpha_{II})(1+\alpha_{EE}) - \alpha_{IE}\alpha_{EI}}\right)  < 1  \label{eq:l1c1}\\
  \lambdaA p^*_S \left(\frac{\beta_1 (1 - \alpha_{II}) + \beta_2 \alpha_{EI}}{(1-\alpha_{II})(1- \alpha_{EE}) - \alpha_{IE}\alpha_{EI}}\right) < 1   \label{eq:l1c2}
 \end{align}
\end{subequations}
\noindent
respectively. The above equations should be true for \textit{any} eigenvalue $\lambdaA$ of $\matA$ which makes $\calD > 0$. Recall that we are considering only undirected graphs, hence $\matA$ is a symmetric binary (0/1) square irreducible matrix. As a result firstly, all its eigenvalues are real. Secondly, from the Perron-Frobenius theorem~\cite{mcculer2000} the algebraically largest eigenvalue $\eig$ of $\matA$ is a positive real number and also has the largest magnitude among all eigenvalues. Hence if the above equations are true for $\lambdaA = \eig$ we are done. 
Now note that \[ (1+ \alpha_{II}) (1+ \alpha_{EE}) - \alpha_{IE}\alpha_{EI} > (1 - \alpha_{II}) (1 - \alpha_{EE}) - \alpha_{IE}\alpha_{EI}\] and that \[ \beta_1 (1 - \alpha_{II}) + \beta_2 \alpha_{EI} > -\beta_1 (1+ \alpha_{II}) + \beta_2 \alpha_{EI}\] In addition the L.H.S in both equations is positive under $\lambdaA = \eig$. So Equation~\ref{eq:l1c1} is always true if Equation~\ref{eq:l1c2} holds (under $\lambdaA = \eig$) i.e.
\begin{equation} 
\eig p^*_S \left(\frac{\beta_1 (1 - \alpha_{II}) + \beta_2 \alpha_{EI}}{(1-\alpha_{II})(1- \alpha_{EE}) - \alpha_{IE}\alpha_{EI}}\right) < 1 
\label{eq:finalrealcond}
\end{equation}

As $\eig$ is the largest eigenvalue both algebraically and in magnitude, under Equation~\ref{eq:finalrealcond},
\begin{eqnarray*}
 \lefteqn{1 - \alpha_{II} \alpha_{EE} + \alpha_{IE}\alpha_{EI} -  p^*_S\lambdaA (\beta_1\alpha_{II}-\beta_2\alpha_{EI})} \\
&> &1 - \alpha_{II} \alpha_{EE} + \alpha_{IE}\alpha_{EI} - \left(\frac{(1-\alpha_{II})(1- \alpha_{EE}) - \alpha_{IE}\alpha_{EI}}{\beta_1 (1 - \alpha_{II}) + \beta_2 \alpha_{EI}}\right) (\beta_1\alpha_{II}-\beta_2\alpha_{EI}) \\
& = & \frac{\left((1-\alpha_{II})^2+\alpha_{IE}\alpha_{EI}\right)\beta_1+\alpha_{EI}(2- \alpha_{II}-\alpha_{EE}) \beta_2}{(1-\alpha_{II})\beta_1 + \alpha_{EI}\beta_2} \\
& > & 0
\end{eqnarray*}
\noindent
$\therefore$ Equation~\ref{eq:qec1} is also true if Equation~\ref{eq:finalrealcond} holds. Thus the condition  for the roots to be in $(-1, 1)$ when they are real is given simply by Equation~\ref{eq:finalrealcond}.
 
\paragraph{$r_1$ and $r_2$ are complex}
In this case $\lambdaA$ is such that $\calD < 0$. Also as Equation~\ref{eq:quad} has real co-efficients, $r_1$ and $r_2$ are complex conjugate of each other and so $ |r_1| = |r_2| = \sqrt{r_1 \cdot r_2} $. But the product of roots $x_1$ and $x_2$ of the equation $a x^2 + bx + c = 0$ is equal to $c/a$. Hence we want to enforce $c/a < 1$. In our case it is \[\alpha_{II} \alpha_{EE} - \alpha_{IE}\alpha_{EI} +  p^*_S\lambdaA (\beta_1\alpha_{II}-\beta_2\alpha_{EI}) < 1\] which is exactly Equation~\ref{eq:qec1}. From the above analysis, we already know that it is true if Equation~\ref{eq:finalrealcond} holds. So, for any eigenvalue $\lambdaA$ for which $\calD < 0$, the roots have magnitude less than 1 given Equation~\ref{eq:finalrealcond} is true.

Thus in both cases, whether roots are real or complex, Equation~\ref{eq:finalrealcond} is a sufficient condition for the roots to have magnitude less than 1. \qed \\

To re-cap we state Theorem~\ref{thm:main} and then give its proof:

\begin{thm}[Super-model theorem - sufficient condition for stability]
For virus propagation models which statisfy our general initial assumptions and for any arbitrary undirected graph with adjacency matrix $\matA$ and largest eigenvalue $\eig$, the sufficient condition for stability is given by:
\begin{equation*}
s  < 1
\end{equation*}
where, $s$ (the \textit{effective strength}) is:
\begin{equation*}
s = \eig \cdot C
\end{equation*}
\noindent
and $C$ is a constant dependent on the model (given by Equation~\ref{eq:c}). Hence, the \tps is reached when $s = 1$.
\end{thm}

\paragraph{Proof}
Lemma~\ref{lem:eigT} and Lemma~\ref{lem:eigquad} ensure cases C1 and C2 and hence together with Lemma~\ref{lem:eigJ} imply that the eigenvalues of the Jacobian $\calJ$ of our general NLDS computed at the fixed point $\vect{x}$ are less than 1 in magnitude if Equation~\ref{eq:finalrealcond} is true. 

$\therefore$ using Theorem~\ref{theorem:asymstable}, our general NLDS is stable at its fixed point $\vect{x}$ if Equation~\ref{eq:finalrealcond} holds. Recall that $\vect{x}$ is the point when there no infected nodes in the system (Appendix~\ref{sec:fixedpt}) and that this is the fixed point whose stability conditions determine the epidemic threshold (Section~\ref{sec:roadmap}). 

$\therefore$ finally we can conclude Theorem~\ref{thm:main} with 
\begin{equation}
\label{eq:c}
C_{\mathrm{\vpm}} =  p^*_S \left(\frac{\beta_1 (1 - \alpha_{II}) + \beta_2 \alpha_{EI}}{(1-\alpha_{II})(1- \alpha_{EE}) - \alpha_{IE}\alpha_{EI}}\right)
\end{equation}
and the effective strength $s = \eig \cdot C_{\mathrm{\vpm}}$. The parameter $C_{\mathrm{\vpm}}$ is a constant for a given propagation model while the only parameter involved from the underlying contact-network is $\eig$, the first eigenvalue of the adjacency matrix. \qed

\end{document}